\documentclass[preprint, 12pt]{elsarticle}

\usepackage{fullpage}
\usepackage{graphicx}
\usepackage{amsmath}
\usepackage{amsthm}
\usepackage{amsfonts}
\usepackage{booktabs}
\usepackage{subfigure}
\usepackage[final]{changes}
\usepackage[super]{nth}

\journal{Computer Physics Communications}

\begin{document}

\newcommand{\td}{\text{d}}
\newcommand{\tD}{\text{D}}
\newcommand{\uvec}[1]{\underline{#1}}
\newcommand{\unabla}{\uvec{\nabla}}
\newcommand{\umat}[1]{\uvec{\uvec{#1}}}
\renewcommand{\emph}[1]{\textit{#1}}

\begin{frontmatter}

\title{Investigation of wall bounded flows using SPH and the unified semi-analytical wall boundary conditions}
\author[manc]{Arno Mayrhofer\corref{cor}}
\ead{arno.m@gmx.at}
\cortext[cor]{Corresponding author}
\author[manc]{Benedict D. Rogers}
\author[edf,lhsv]{Damien Violeau}
\author[edf]{Martin Ferrand}

\address[manc]{School of Mechanical Aerospace and Civil Engineering, University of Manchester, Manchester, United Kingdom}
\address[edf]{EDF R\&D, Chatou, France}
\address[lhsv]{Saint-Venant Laboratory for Hydraulics, Universit\'e Paris-Est, Paris, France}

\begin{abstract}
The semi-analytical wall boundary conditions present a mathematically rigorous framework to prescribe the influence of solid walls in SPH for fluid flows. In this paper they are investigated with respect to the skew-adjoint property which implies exact energy conservation. It will be shown that this property holds only in the limit of the continuous SPH approximation, whereas in the discrete SPH formulation it is only approximately true, leading to numerical noise. This noise, interpreted as form of ``turbulence'', is treated using an additional volume diffusion term in the continuity equation which we show is equivalent to an approximate Riemann solver. Subsequently two extensions to the boundary conditions are presented. The first dealing with a variable driving force when imposing a volume flux in a periodic flow and the second showing a generalization of the wall boundary condition to Robin type and arbitrary-order interpolation. Two modifications for free-surface flows are presented for the volume diffusion term as well as the wall boundary condition. In order to validate the theoretical constructs numerical experiments are performed showing that the present volume flux term yields results with an error 5 orders of magnitude smaller then previous methods while the Robin boundary conditions are imposed correctly with an error depending on the order of the approximation. Furthermore, the proposed modifications for free-surface flows improve the behaviour at the intersection of free surface and wall as well as prevent free-surface detachment when using the volume diffusion term. Finally, this paper is concluded by a simulation of a dam break over a wedge demonstrating the improvements proposed in this paper.
\end{abstract}


\end{frontmatter}

\section{Introduction}
\label{s:int}
Smoothed particle hydrodynamics (SPH) has become one of the predominant meshless methods to solve the Navier-Stokes equations. An introduction to this method can be found in the reference paper by Monaghan \cite{monaghan_smoothed_2005} or the book by Violeau \cite{violeau_fluid_2012}. One of the greatest difficulties associated with this method is the imposition of boundary conditions. The semi-analytical wall boundary conditions introduced by Ferrand \emph{et al.} \cite{ferrand_unified_2012} provide a consistent framework for imposing wall boundary conditions in both laminar and turbulent flows. Contrary to previous approaches such as the Lennard-Jones potential forces (see Monaghan \cite{monaghan_simulating_1994}) or more advanced repulsive force models (see Monaghan and Kajtar \cite{monaghan_sph_2009}) they do not rely on additional modelling but instead are introduced naturally from the numerical treatment of the governing equations. Furthermore, they can deal with arbitrarily complex geometries in contrast to the approach using ghost particles (see \emph{e.g.} Colagrossi and Landrini \cite{colagrossi_numerical_2003}).\\
Despite its ability to predict flows with good accuracy, SPH with the semi-analytical wall boundary conditions still suffers from numerical drawbacks. This can be observed, for example, when considering the energy budget from an inviscid flow between two infinite plates. As the energy is not constant the skew-adjoint principle is violated thus warranting an investigation of this property with respect to the semi-analytical wall boundary conditions. In addition to a theoretical analysis a numerical simulation will show stability issues associated with this property. This leads to the question of whether it is possible to counter the instabilities by some numerical procedure. If a diffusion term is chosen, as in this paper, then its properties and influence on the fluid flow need to be investigated. The question of how to impose a volume flux or specific boundary conditions are also closely linked to modelling wall boundary conditions. In free-surface flows certain difficulties can occur which are not apparent in confined flows. To solve these, it is necessary to revisit certain investigations by separating hydrostatic and dynamic pressure differences. \added{Within the development of SPH formulations, numerous modifications have been suggested to improve the behaviour of SPH.  In this paper, we unify these modifications (energy conservation, diffusion correction, Ferrand \emph{et al.} \cite{ferrand_unified_2012} treatment, etc.), investigate their properties and extend them to more general situations.}
\\
In the following section the wall boundary conditions will be introduced in detail and will then be investigated with respect to the skew-adjoint property in Section \ref{s:sa}. This will be done by considering analytical as well as numerical calculations. The instabilities that will be observed will then be treated by using a volume diffusion term that is derived from turbulent Reynolds-averaged considerations and implemented into the continuity equation. This term is related to an approximate Riemann solver proposed by Ferrari \emph{et al.} \cite{ferrari_new_2009} and adapted to the present wall boundary conditions.\\
The next two sections deal with improvements regarding the imposition of boundary values. In Section \ref{s:volflux} a new formula for imposing a non-constant driving force based on a volume flux will be introduced and compared to a standard formulation. Subsequently, the method of imposing Neumann wall boundary conditions will be generalized to Robin boundary conditions and arbitrary orders of accuracy. A wave equation with Robin boundary conditions will be solved numerically in order to demonstrate the capability of the present model.\\
The last theoretical contribution presented in this paper will then deal with minor modifications to the volume diffusion term as well as the wall boundary conditions to take external forces, \emph{e.g.} gravity, into account. These two modifications are validated by two still water simulations. \\
Finally, the present methodology is applied to a violent free-surface flow, a schematic dam-break over a wedge. Quantitative comparison of wall pressure forces is made with respect to other numerical formulations including a Volume-of-Fluid method and it is shown that the volume diffusion term avoids using any free-surface correction in the continuity equation required in previous work \cite{ferrand_unified_2012}.

\section{SPH Overview}
\label{s:overview}
\begin{figure}[htb]
\centering
\includegraphics[width=0.40\textwidth]{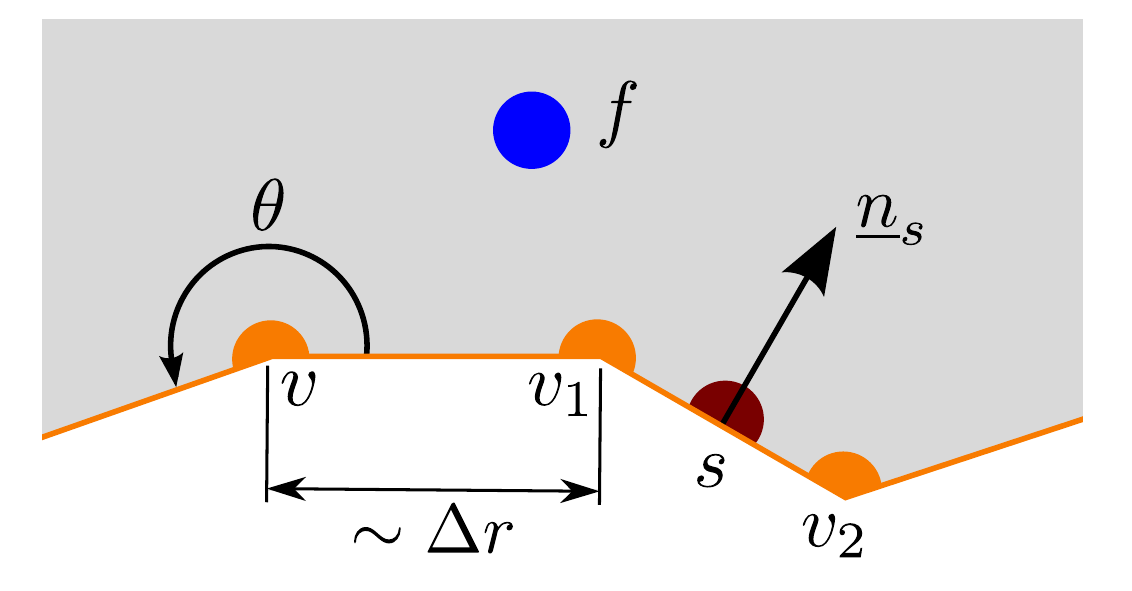}
\caption{The different types of elements.}
\label{f:int:elements}
\end{figure}

Smoothed particle hydrodynamics (SPH) is now a well-known Lagrangian numerical method. It will be assumed that the reader is familiar with the basics (for theoretical grounds refer to Monaghan \cite{monaghan_simulating_1994} or Violeau \cite{violeau_fluid_2012}). Particles will be denoted by $a$ or $b$, and the interpolating SPH kernel by $w$. It is assumed to be radial, \textit{i.e.} isotropic. The kernel used throughout the paper is the \nth{5}-order Wendland kernel given by
\begin{equation}
	w(r) = \left\{ \begin{array}{cl}
		\dfrac{\alpha_m}{h^m}\left(1-\dfrac{1}{2}\dfrac{r}{h}\right)^4\left(1+2\dfrac{r}{h}\right) & \mbox{ if } 0 \le r/h \le 2\mbox{,}\\
		0 & \mbox{ if } 2 < r/h\mbox{,}
	\end{array} \right.
\end{equation}
where $r$ is the distance between two particles, $m$ is the dimension, $\alpha_m = 7/(4\pi)$ in 2-D and $h$ is the smoothing length. The latter in the following is set as $h= 2\Delta r$, where $\Delta r$ is the initial particle spacing. Only weakly compressible SPH \cite{monaghan_smoothed_1992} in 2-D will be considered in this paper.\\
Semi-analytical wall boundary conditions for arbitrary boundaries in SPH were introduced by Ferrand \textit{et al.} in \cite{ferrand_unified_2012,ferrand_improved_2010,ferrand_consistent_2011}. The approach is inspired by the papers from Kulasegaram \textit{et al.} \cite{kulasegaram_variational_2004} and Di Monaco \textit{et al.} \cite{di_monaco_sph_2011}. First, the main principles of this method are recalled. The key idea is to use an analytical kernel wall correction factor $\gamma$ defined by
\begin{equation}
\gamma_a = \int_\Omega w_{ab} \uvec{\td r}_b,
\label{e:overview:gam}
\end{equation}
where $w_{ab} = w(\|\uvec{r}_a - \uvec{r}_b\|)$, $\uvec{r}_a$  being the position of particle $a$. Here, $\uvec{r}_b$ is used as a continuous variable, integrated over the fluid domain $\Omega$. \added[remark= Reviewer 1: Remark 1]{If no part of the wall is inside the kernel support of a particle $a$ then the value of $\gamma_a$ is equal to one.}\\
The boundary conditions are implemented by using the sets of elements described in Table \ref{t:int:elements}.
As usual, the fluid is discretized using particles with a spacing initially set to $\Delta r$. The boundary is split into line segments $s$ of length approximately $\Delta r$. These segments are called boundary elements and are located between two vertex particles $v$ as shown in Figure \ref{f:int:elements}.
\begin{table}[htb]
\centering
\begin{tabular}{c|p{0.40\textwidth}}
\textbf{Set} & \textbf{Description} \\
\midrule
$\mathcal{P}$ & All particles ($\mathcal{V} \cup \mathcal{F}$)\\
$\mathcal{V}$ & Particles on wall boundary (vertex particles)\\
$\mathcal{F}$ & Particles inside the fluid domain\\
$\mathcal{S}$ & Wall boundary segments
\end{tabular}
\caption[Overview of elements]{Overview of elements.}
\label{t:int:elements}
\end{table}\\
Herein, differential operators, both continuous and discrete, will be written in bold (\textit{e.g.} $\textbf{Div}$). Vectors and matrices will be written as $\uvec{B}$ and $\umat{M}$ respectively.
The general discrete SPH approximation for a scalar $f$ at position $a$ is given by
\begin{equation}
[f]_a = \frac{1}{\gamma_a} \underset{b\in\mathcal{P}}{\sum}V_b f_b w_{ab},
\label{e:int:sph-approx}
\end{equation}
where $V_b$ is the volume of particle $b$, $\mathcal{P}$ the set of all particles and $f_b$ the value of the field $f$ associated with particle $b$. The continuous analog will be denoted with $[f]^c_a$. Note that in general $[f]_a^{(c)} \neq f_a$, \textit{i.e.} the SPH interpolation violates the Kronecker delta property even in the continuous case.\\
To obtain an approximation of $\gamma$, a governing equation is used, given by
\begin{equation}
\frac{\td\gamma_a}{\td t} = \uvec{v}^r_a\cdot\unabla\gamma_a,
\label{e:int:gam-gov}
\end{equation}
where $\uvec{v}^r$ is the velocity relative to the wall and 
\begin{equation}
\unabla \gamma_a = \int_{\partial \Omega} w_{ab} \uvec{n}_b \uvec{\td r}_b.
\end{equation}
The normal $\uvec{n}_b$ in the equation above is oriented inwards, a convention used throughout this paper. Note that the integral in Eq. \eqref{e:overview:gam} is obtained by a discretization in time and values based on the boundary. It is thus distinctly different to the Shepard filter given as
\begin{equation}
\alpha_a = \underset{b\in\mathcal{P}}{\sum} V_b w_{ab},
\label{e:overview:sheppard}
\end{equation}
which originates from a discretization in space based on the fluid particles.\\
The other key idea here is to derive rigorously the differential operators without neglecting boundary terms coming from the integration by parts as demonstrated below starting from the continuous SPH approximation (denoted by the superscript $c$) of a gradient of a scalar field $f$:
\begin{eqnarray}
[\unabla f]^c_a & = & \frac{1}{\gamma_a}\int_\Omega (\unabla f)_b w_{ab} \uvec{\td r}_b
\label{e:overview:intbyparts}\\
& = & - \frac{1}{\gamma_a}\int_\Omega f_b \unabla_b w_{ab} \uvec{\td r}_b + \frac{1}{\gamma_a}\int_\Omega \unabla_b(f_b w_{ab}) \uvec{\td r}_b\nonumber\\
& = & \frac{1}{\gamma_a}\int_\Omega f_b \unabla_a w_{ab} \uvec{\td r}_b - \frac{1}{\gamma_a}\int_{\partial \Omega} f_b w_{ab} \uvec{n}_b \uvec{\td r}_b.\nonumber
\label{e:overview:green}
\end{eqnarray}
To obtain the last line the kernel gradient asymmetry (a result of kernel isotropy) was used as well as Green's Theorem. From the SPH approximation as given by Eq. \eqref{e:int:sph-approx} Ferrand \textit{et al.} proposed the following SPH approximation for the divergence of a vector field $\uvec{B}$:
\begin{equation}
\textbf{Div}^{\gamma,F}_a(\uvec{B}) := - \frac{1}{\gamma_a \rho_a}\underset{b\in\mathcal{P}}{\sum} m_b \uvec{B}_{ab} \cdot \unabla_a w_{ab} + \frac{1}{\gamma_a}\underset{s\in\mathcal{S}}{\sum}\uvec{B}_{as} \cdot \unabla \gamma_{as},
\label{e:int:div-orig}
\end{equation}
where the superscript $F$ stands for "Ferrand" and $\uvec{B}_{ab} = \uvec{B}_a-\uvec{B}_b$ and
\begin{equation}
\unabla \gamma_{as} = \int_s w_{ab}\uvec{n}_b \uvec{\td r}_b,
\label{e:int:ggamma}
\end{equation}
is the integral of the kernel over an individual segment $s$, \textit{i.e.} the contribution of this segment to $\unabla \gamma_a$. Hence, the gradient of $\gamma_a$ can be written as
\begin{equation}
\unabla \gamma_a = \underset{s\in\mathcal{S}}{\sum} \unabla \gamma_{as}.
\end{equation}
Note that the integral in Eq. \eqref{e:int:ggamma} is known analytically for a given kernel (see Ferrand \textit{et al.} \cite{ferrand_unified_2012} for the case of the Wendland kernel). Following the same paper, the discrete gradient of a scalar field $f$ is given by
\begin{equation}
\textbf{Grad}^{\gamma,F}_a(f) = \frac{\rho_a}{\gamma_a}\underset{b\in\mathcal{P}}{\sum}m_b\left(\frac{f_a}{\rho_a^2}+\frac{f_b}{\rho_b^2}\right)\unabla_a w_{ab}
 - \frac{\rho_a}{\gamma_a}\underset{s\in\mathcal{S}}{\sum}\left(\frac{f_a}{\rho_a^2}+\frac{f_s}{\rho_s^2}\right)\rho_s\unabla \gamma_{as}.
\label{e:int:grad-orig}
\end{equation}
The boundary terms in the above differential operators originate from the fact that the differential operator is shifted from the unknown function to the kernel via partial integration. In most SPH publications they are neglected. This is not the case in the semi-analytical wall boundary conditions where they are transformed to boundary integrals via Green's theorem as shown in Eq. \eqref{e:overview:green}.\\
The continuity equation used in \cite{ferrand_unified_2012} is not the conventional time-dependent form given as
\begin{equation}
\frac{\td \rho_a}{\td t} = - \rho_a \textbf{Div}^{\gamma,F}_a(\uvec{v}),
\label{e:overview:td-cont-eq}
\end{equation}
but the equivalent time-independent form
\begin{equation}
\td (\gamma_a \rho_a) = \td \left(\underset{b\in\mathcal{P}}{\sum} m_b w_{ab}\right).
\label{e:overview:ti-cont-eq}
\end{equation}
In the case of a free-surface flow $\gamma_a$ is replaced with an heuristic blending function which is a necessary correction as otherwise particle repulsion can be observed as demonstrated later.\\
The Navier-Stokes equation is given as
\begin{equation}
\frac{\td \uvec{v}_a}{\td t} = - \frac{1}{\rho_a}\textbf{Grad}^{\gamma,F}_a(p) + \textbf{Lap}^{\gamma,F}_a(\mu, \uvec{v}) + \uvec{g},
\label{e:overview:ns}
\end{equation}
where the SPH Laplacian is defined below by Eq. \eqref{e:sph:lap}. The system of equations is closed with the Tait equation of state \cite{tait_report_1888}, given as
\begin{equation}
p = \frac{\rho_0 c_0^2}{\zeta}\left[\left(\frac{\rho}{\rho_0}\right)^\zeta-1\right],
\label{e:overview:eos}
\end{equation}
where $\rho_0$ is the reference density, $c_0$ the numerical speed of sound and $\zeta = 7$ for water.\\
Continuing to follow \cite{ferrand_unified_2012}, in order to impose Neumann boundary conditions at the wall for a scalar $f$ the values from the fluid can be extrapolated to a vertex particle $v$ via
\begin{equation}
f_v = \frac{1}{\alpha_v}\underset{b\in\mathcal{F}}{\sum}V_b\;f_b\;w_{vb},
\label{e:overview:neumann-bcs}
\end{equation}
where
\begin{equation}
\alpha_v = \underset{b\in\mathcal{F}}{\sum}V_b\;w_{bv}.
\label{e:overview:alpha-v}
\end{equation}
The values for the boundary segments $s$ are given as an arithmetic mean of the associated vertex particles. The above represents a first-order approximation of the true Neumann condition.\\
Finally, some aspects of the viscosity treatment as proposed in \cite{ferrand_unified_2012} will be summarized. The first requirement is a Laplacian operator for which the same approach as Morris \textit{et al.} \cite{morris_modeling_1997} is used but including the boundary terms. Ferrand \textit{et al.} \cite{ferrand_unified_2012} approximate the Laplacian as
\begin{equation}
\unabla\cdot(f\unabla\otimes\uvec{B}) \approx \textbf{Lap}^{\gamma,F}_a(f,\uvec{B}) = \frac{\rho_a}{\gamma_a}\underset{b\in\mathcal{P}}{\sum}m_b\frac{f_a+f_b}{\rho_a\rho_b}\frac{\uvec{B}_{ab}}{r_{ab}^2}\uvec{r}_{ab}\cdot\unabla_a w_{ab} - \frac{2}{\gamma_a}\underset{s\in\mathcal{S}}{\sum}f_s (\unabla \uvec{B})_s \cdot \unabla \gamma_{as},
\label{e:sph:lap}
\end{equation}
where $r_{ab} = \|\uvec{r}_{ab}\|$. In the case of viscosity $\uvec{B} = \uvec{v}, f=\mu$ and so $\mu_s(\unabla\otimes\uvec{v})_s\cdot\uvec{n}_s = \uvec{\tau}_s$, \textit{i.e.} the wall shear stress vector at segment $s$. The latter is given by
\begin{equation}
\mu\frac{\partial\uvec{v}}{\partial\uvec{n}} = \rho \|\uvec{v}_\tau\|\uvec{v}_\tau,
\end{equation}
where, by definition, $\uvec{v}_\tau$ is the friction velocity. In the case of laminar flow this is approximated by
\begin{equation}
\|\uvec{v}_\tau\|\uvec{v}_\tau = \frac{\nu \uvec{v}(z)}{z},
\end{equation}
where $z$ is a short distance from the wall and $\nu=\mu/\rho$. This equation only holds for $z \ll L$, where $L$ is a characteristic length-scale of the flow geometry. It can be extended to turbulent flows using wall functions, see \cite{ferrand_unified_2012}.

\section{On the skew-adjoint property including boundaries}
\label{s:sa}
\subsection{Theoretical investigation}
Now, new developments will be presented in order to better understand and improve Ferrand \emph{et al.}'s \cite{ferrand_unified_2012} model. \added{The boundary corrected formulation has already been theoretically investigated by Macia \emph{et al.} \cite{macia_boundary_2012} with respect to its approximation property in 1-D. The goal of this section is to investigate another aspect of this formulation which is the energy conservation.}
\\
In this section the focus will be on the property of skew-adjointness of the two arbitrary (discrete or continuous) operators $\textbf{Grad}$ and $\textbf{Div}$. These two operators are skew-adjoint if and only if
\begin{equation}
<\textbf{Grad}(f),\uvec{B}> = - <f,\textbf{Div}(\uvec{B})>,
\label{e:sa:def}
\end{equation}
where $\left<\right>$ are the respective $L^2$ scalar products. Before starting the actual investigation the importance of this property shall be highlighted. Consider a system of particles representing a fluid without external influence nor dissipative forces, then its energy is given by
\begin{equation}
E = E_{kin} + E_{int} = \underset{a\in\mathcal{P}}{\sum}m_a\left( \frac{1}{2}\|\uvec{v}_a\|^2 + e_{int,a}(\rho_a)\right),
\end{equation}
where $E_{kin}$ and $E_{int}$ are the total kinetic and internal energy respectively, while $e_{int,a}$ is the specific internal energy of particle $a$. The time derivative of the kinetic energy is given by
\begin{equation}
\frac{\td E_{kin}}{\td t} = \underset{a\in\mathcal{P}}{\sum} m_a \uvec{v}_a\cdot \frac{\td \uvec{v}_a}{\td t} = -\underset{a\in\mathcal{P}}{\sum} m_a \uvec{v}_a\cdot \frac{1}{\rho_a}\textbf{Grad}_a(p),
\end{equation}
where the Euler momentum equation was used, \emph{i.e.} Eq. \eqref{e:overview:ns} without the $\textbf{Lap}$ and $\uvec{g}$ terms. The time derivative of the internal energy can in turn be written as
\begin{equation}
\frac{\td E_{int}}{\td t} = \underset{a\in\mathcal{P}}{\sum} m_a \left(\frac{\partial e_{int}}{\partial \rho}\right)_a\frac{\td \rho_a}{\td t} = \underset{a\in\mathcal{P}}{\sum} \frac{m_a p_a}{\rho_a^2}\frac{\td \rho_a}{\td t},
\end{equation}
where the last equality follows from thermodynamic principles that relate the internal energy per unit mass to pressure and density via $p = \rho^2\,\td e_{int}/\td \rho$. Looking at the time derivative of the total energy one thus obtains with the help of the continuity equation (Eq. \eqref{e:overview:td-cont-eq}):
\begin{equation}
\frac{\td E}{\td t} = - \underset{a \in \mathcal{P}}{\sum} V_a\uvec{v}_a\cdot \textbf{Grad}_a(p) - \underset{a \in \mathcal{P}}{\sum} V_a p_a \textbf{Div}_a(\uvec{v}).
\end{equation}
Written in notation with discrete scalar products this yields
\begin{equation}
\frac{\td E}{\td t} = - \left<\textbf{Grad}_a(p), \uvec{v}_a\right> - \left< p_a, \textbf{Div}_a(\uvec{v})\right>.
\label{e:sa:dEdt}
\end{equation}
This shows that the energy is exactly conserved if the two discrete differential operators $\textbf{Grad}$ and $\textbf{Div}$ are skew-adjoint (Eq. \eqref{e:sa:def}), \emph{i.e.} if the right-hand-side of \eqref{e:sa:dEdt} is equal to zero. This property is natural, since the same occurs with ordinary differential operators when changing the discrete sums with integrals (see \emph{e.g.} Violeau \cite{violeau_fluid_2012}):
\begin{equation}
\mbox{Skew-adjoint operator definition:}\qquad\underbrace{\left<\unabla f,\uvec{B}\right> + \left<f,\unabla\cdot\uvec{B}\right>}_{=:SA} = -\int_{\partial\Omega}f(\uvec{r})\,\uvec{B}(\uvec{r})\cdot\uvec{n}(\uvec{r}) \td r.
\label{e:sa:sa-ana}
\end{equation}
If $f=p$ and $\uvec{B} = \uvec{v}$, the pressure and velocity, respectively, which solve the Navier-Stokes equations, then the right-hand side is equal to zero. This is due to $p=0$ at a free-surface and $\uvec{v}\cdot\uvec{n} = 0$ at a solid wall. It would thus be advantageous if the SPH operators adhere to this property.\\
The investigation \replaced{into}{inot} skew-adjoint operators will commence by first considering basic continuous SPH operators without boundary terms, \emph{i.e.} using integrals instead of discrete sums and assuming that no boundaries are present. The operators under investigation are given by
\begin{eqnarray}
\textbf{Grad}^{b,c}_a(f) & = & \int_\Omega f_b \unabla_a w_{ab} \uvec{\td r}_b\\
\textbf{Div}^{b,c}_a(\uvec{B}) & = & \int_\Omega \uvec{B}_b\cdot \unabla_a w_{ab} \uvec{\td r}_b,
\end{eqnarray}
where the superscript $b$ stands for ``basic'' in the sense that it is the SPH operator which can be derived directly from the interpolation without the addition of any other terms. The superscript $c$ stands for ``continuous'' similar to the SPH interpolation. The left-hand side of the skew-adjoint operator definition (Eq. \eqref{e:sa:sa-ana}) is then given by
\begin{eqnarray}
SA = \int_\Omega\int_\Omega\left( f_b\,\unabla_a w_{ab} \cdot \uvec{B}_a + f_a \,\uvec{B}_b\cdot \unabla_a w_{ab}\right) \uvec{\td r}_b\uvec{\td r}_a.
\end{eqnarray}
Due to integral additivity it is possible to swap the dummy labels $a,b$ in the second term which yields
\begin{eqnarray}
SA = \int_\Omega\int_\Omega\left( f_b\,\uvec{B}_a \cdot \unabla_a w_{ab} + f_b \,\uvec{B}_a\cdot \unabla_b w_{ab}\right) \uvec{\td r}_b\uvec{\td r}_a.
\end{eqnarray}
The kernel isotropy result in the asymmetry of its gradient, \emph{i.e.} $\unabla_a w_{ab} = - \unabla_b w_{ab}$, shows finally that $SA = 0$, \emph{i.e.} that $\textbf{Grad}^{b,c}$ and $\textbf{Div}^{b,c}$ are skew-adjoint. Note that this property also holds true in the discrete case, where the integrals are replaced by sums, from the same properties.\\
Now the following symmetrised operators shall be defined:
\begin{eqnarray}
\textbf{Grad}^{s,c,k}_a(f) & = & \int_\Omega \frac{\rho_a^{2k}f_b+\rho_b^{2k}f_a}{\rho_a^k\rho_b^k} \unabla_a w_{ab} \uvec{\td r}_b, \label{e:sa:grad-s}\\
\textbf{Div}^{s,c,k}_a(\uvec{B}) & = & \frac{1}{\rho_a^{2k}}\int_\Omega \rho_a^k\rho_b^k(\uvec{B}_b-\uvec{B}_a)\cdot \unabla_a w_{ab} \uvec{\td r}_b \label{e:sa:div-s},
\end{eqnarray}
where the superscript $s$ stands for ``standard'' and $k$ is a power \replaced{used to discuss a wide range of operators using a sole notation. Its effect will be discussed in the following}{to be discussed}. Note that, apart from the lack of boundary terms, the symmetrized gradient \eqref{e:sa:grad-s} differs from the discrete form \eqref{e:int:grad-orig} by the presence of density terms. Eq. \eqref{e:int:grad-orig} can be recovered by setting $k=1$. The above operators in Equations \eqref{e:sa:grad-s} and \eqref{e:sa:div-s} are skew-adjoint if the newly added terms have opposing signs. The proof, which will be omitted, can be found in \cite{violeau_fluid_2012}. As in the previous case it holds true in the discrete case also.\\
When removing the assumption of no boundaries the calculations are no longer as straightforward. Following a similar procedure to Equation \eqref{e:overview:intbyparts}, in the vicinity of a boundary the operators of interest are given in continuous form by
\begin{eqnarray}
\textbf{Grad}^{\gamma,c,k}_a(f) & = & \frac{1}{\gamma_a}\int_\Omega \frac{\rho_a^{2k}f_b+\rho_b^{2k}f_a}{\rho_a^k\rho_b^k} \unabla_a w_{ab} \uvec{\td r}_b
\label{e:sa:gradcont}\\
&-&\frac{1}{\gamma_a}\int_{\partial\Omega} \frac{\rho_a^{2k}f_b+\rho_b^{2k}f_a}{\rho_a^k\rho_b^k} \uvec{n}_b w_{ab} \uvec{\td r}_b,\nonumber\\
\textbf{Div}^{\gamma,c,k}_a(\uvec{B}) & = & \frac{1}{\gamma_a \rho_a^{2k}}\int_\Omega \rho_a^k\rho_b^k(\uvec{B}_b-\uvec{B}_a)\cdot \unabla_a w_{ab} \uvec{\td r}_b
\label{e:sa:divcont}\\
&-&\frac{1}{\gamma_a\rho_a^{2k}}\int_{\partial\Omega} \rho_a^k\rho_b^k(\uvec{B}_b - \uvec{B}_a)\cdot\uvec{n}_b w_{ab} \uvec{\td r}_b\nonumber,
\end{eqnarray}
where the superscript $\gamma$ indicates the renormalization. Apart from the additional density terms in the denominator of the $\textbf{Grad}$ operator, these formulae are the continuous analog to Eqs. \eqref{e:int:div-orig} and \eqref{e:int:grad-orig} for $k=1$. In \ref{s:sa-app} using the gradient and divergence given by Eqs. \eqref{e:sa:gradcont} and \eqref{e:sa:divcont} the left-hand side of the skew-adjoint operator definition, Eq. \eqref{e:sa:sa-ana}, can be shown to reduce to \deleted{. The final result is given by}
\begin{equation}
SA \underset{(h\rightarrow0)}{\rightarrow} - \int_{\partial \Omega}f_a\uvec{B}_a\cdot\uvec{n}_a \uvec{\td r}_a.
\end{equation}
This is equivalent to Eq. \eqref{e:sa:sa-ana} showing the skew-adjoint property for SPH continuous operators with boundary terms in case of the correct imposition of the boundary conditions in the limit $h\rightarrow0$. Note that it is essential for this result that the operators are renormalized with $\gamma$. Furthermore, the process of taking the limit is only required due to the violation of the Kronecker delta property by the SPH interpolation.\\
The derivation shown in \ref{s:sa-app} could have been made significantly shorter, if we were only interested in continuous operators. However, the reason for explicitly going through this derivation was to show the steps necessary to prove the same in the case of discrete operators. It would thus be necessary to have a discrete version of the Stokes' theorem. Additionally, the Kronecker delta property is required, but that is violated even by the continuous SPH interpolation. The difficulty for a discrete Stokes' theorem is due to the fact that $\unabla \gamma$ is calculated analytically along the wall (see Section \ref{s:overview}), whereas the volumetric integral over $\unabla w$ is approximated via a discrete sum.\\
As a consequence, discrete SPH operators with boundary terms as presented here are not exactly skew-adjoint contrary to continuous SPH operators. Inside the fluid, however, ($\gamma = 1$, no boundary terms) the skew-adjoint property is fulfilled in the discrete case as it is equivalent to the standard differential operators ($\textbf{Div}^{s,c,k}$, $\textbf{Grad}^{s,c,k}$) given by Eqs. \eqref{e:sa:grad-s} and \eqref{e:sa:div-s}.\\
Finally, the discretized forms of Eqs. \eqref{e:sa:gradcont} and \eqref{e:sa:divcont} are given by
\begin{eqnarray}
\textbf{Grad}^{\gamma,k}_a(f) & = & \frac{1}{\gamma_a}\underset{b\in\mathcal{P}}{\sum}V_b \frac{\rho_a^{2k}f_b+\rho_b^{2k}f_a}{\rho_a^k\rho_b^k} \unabla_a w_{ab} -\frac{1}{\gamma_a}\underset{s\in\mathcal{S}}{\sum} \frac{\rho_a^{2k}f_s+\rho_s^{2k}f_a}{\rho_a^k\rho_s^k}\unabla\gamma_{as}
\label{e:sa:grad-sa}\\
\textbf{Div}^{\gamma,k}_a(\uvec{B}) & = & \frac{1}{\gamma_a\rho_a^{2k}} \underset{b\in\mathcal{P}}{\sum}V_b \rho_a^k\rho_b^k(\uvec{B}_b-\uvec{B}_a)\cdot \unabla_a w_{ab} -\frac{1}{\gamma_a\rho_a^{2k}}\underset{s\in\mathcal{S}}{\sum} \rho_a^k\rho_s^k(\uvec{B}_s - \uvec{B}_a)\cdot\unabla \gamma_{as}.
\label{e:sa:div-sa}
\end{eqnarray}
Note that these operators are a generalization from the ones derived by Ferrand \emph{et al.} given by Eqs. \eqref{e:int:div-orig} and \eqref{e:int:grad-orig} (where $k=1$). In the following $k=0$ will be used unless otherwise noted. According to the authors experience this choice has at best a marginal influence on the results.

\subsection{A numerical experiment}
\label{s:sa:num}
As the above theoretical investigation indicates the SPH discrete operators for the semi-analytical boundary conditions are not skew-adjoint, and thus do not conserve total energy. Hence, a flow between \replaced{two}{to} infinite plates \added{(at $z=0$ and $z=h$)} will be used to address this issue from a numerical perspective. Equations \eqref{e:overview:td-cont-eq} and \eqref{e:overview:ns} are solved with the pressure calculated via the equation of state \eqref{e:overview:eos} and the particles are moved according to $\td \uvec{r}_a/\td t = \uvec{v}_a$. The viscosity is set to zero and the following velocity profile is used as initial condition
\begin{equation}
\added{v_x = 0,} v_z = \frac{c_0}{10}\sin(4\pi z), \nonumber
\end{equation}
where $c_0$  is the numerical speed of sound and $z\in[0,1]$.
The quantity of interest is the energy $E$ which is defined by
\begin{equation}
E(t) = \underset{n=1}{\overset{t}{\sum}}\left(<\textbf{Grad}^{\gamma}(p),\uvec{v}> + <p,\textbf{Div}^{\gamma}(\uvec{v})>\right)^n,
\label{e:sa:energy}
\end{equation}
where the subscript $n$ indicates the current time-step. The value of $E$ is equal to the time integral of the left hand side of Eq. \eqref{e:sa:sa-ana} and it is zero if the operators under investigation are skew-adjoint.
\begin{figure}[htb]
	\centering
	\includegraphics[width=0.70\textwidth]{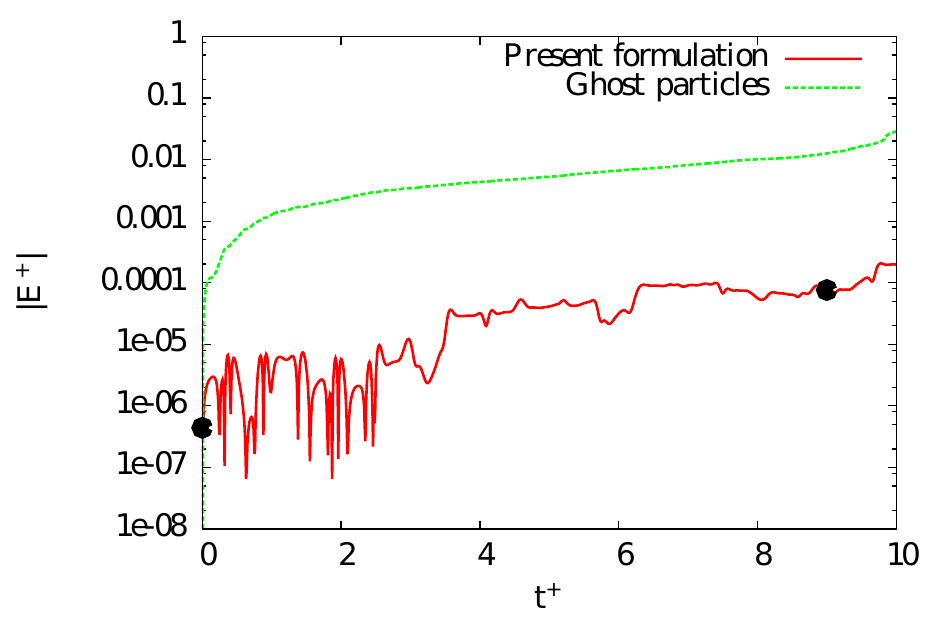}
	\caption{Non-dimensional energy budget over time $t^+=tc_0/h$ for ghost particles and present approach}
	\label{f:sa:num:sa}
\end{figure}
\\
In Figure \ref{f:sa:num:sa} the non-dimensionalized value $E^+ = E/(\rho_0 c_0^2 h^2)$ can be seen plotted over time for the present boundary condition as well as the ghost particle approach \cite{libersky_high_1993} \added{using Eqs. \eqref{e:sa:grad-s} and \eqref{e:sa:div-s} for the differential operators}. Two important features of this plot shall be highlighted. The first is that the value $E$ is never zero, indicating that the operators are indeed not analytically skew-adjoint in cases where boundaries are present. Compared to the ghost particle formulation the energy introduced to the flow is lower with the present boundary conditions.\\
It shall be noted that this simulation was also run with the formulation by Kulasegaram \cite{kulasegaram_variational_2004} which resembles the present approach but, in contrast to the present method, is analytically skew-adjoint. However, as Ferrand \cite{ferrand_unified_2012} already pointed out, the operators by Kulasegaram lack a term in the gradient and thus do not represent the physics accurately. The simulation was also conducted with the formulation by Monaghan and Kajtar \cite{monaghan_sph_2009} but the results were completely non-physical.
\begin{figure}[htb]
\centering
\subfigure[]{
	\includegraphics[width=0.46\textwidth]{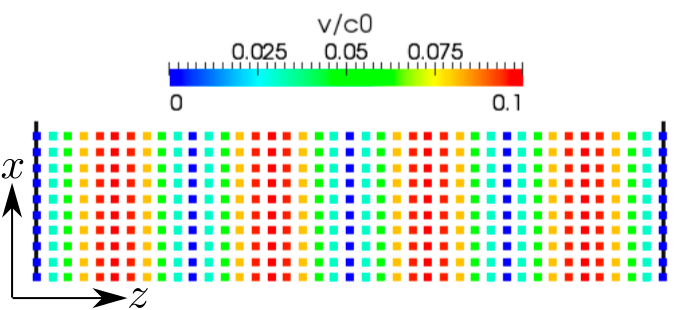}
	\label{f:sa:num:v-init}
}
\subfigure[]{
	\includegraphics[width=0.46\textwidth]{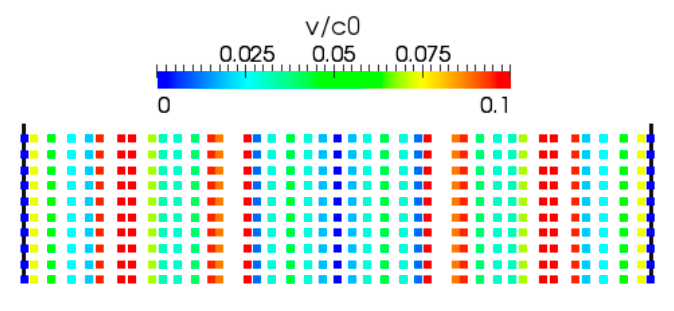}
	\label{f:sa:num:v-spurious}
}
\caption{Values of $v^+$ at different times ($t^+=0.0$: Fig. \ref{f:sa:num:v-init}, $t^+=9.0$: Fig. \ref{f:sa:num:v-spurious})}
\label{f:sa:num:flow}
\end{figure}
\\
The second observation is that the flow starts exhibiting non-physical fluctuations after some time. In Figure \ref{f:sa:num:flow} the flow can be seen at $t^+=t\,c_0/h=0$ and $t^+=9.0$ (corresponding to the times indicated by black circles in Figure \ref{f:sa:num:sa}) where the non-dimensionalized values of $v$ is plotted at each particle. It was already remarked in the previous section that the deficiency has to originate from the boundaries, as inside the fluid the operators are skew-adjoint.\\
This result can be connected to the numerical experiments of Basa \emph{et al.} \cite{basa_robustness_2009} who simulate the laminar Poiseuille flow with viscosity. They also observe that the break-up that occurs, originates from the boundaries as particles which are close to the wall experience different viscous forces due to numerical fluctuations. It should be noted that the behaviour of the viscous system is quite similar to that of a laminar flow that becomes turbulent. Starting from this observation the continuity equation will be investigated in the next section to extract the mean flow field in order to suppress this numerical ``turbulence''.

\section{A volume diffusion term for numerical turbulence}
\label{s:vdt}
\subsection{Basic idea}
\label{s:vdt:idea}
In the context of turbulent flows the continuity equation in the Reynolds averaged context is given by
\begin{equation}
\frac{\overline{\td} \overline{\rho}}{\overline{\td} t} = - \overline{\rho}\unabla\cdot\overline{\uvec{v}} - \unabla\cdot(\overline{\rho'\,\uvec{v}'}),
\label{e:vdt:rans-cont-eq}
\end{equation}
where the primes refer to turbulent fluctuations and the bars the Reynolds averaging (see \emph{e.g.} Pope \cite{pope_turbulent_2001}). Moreover, the bar on the Lagrangian derivative means that the advection velocity is in principle the averaged one. In laminar flows, as considered here, the fluctuating quantities are supposed to be zero. Still Eq. \eqref{e:vdt:rans-cont-eq} can be applied to numerical fluctuations in an attempt to stabilize them. The gradient-diffusion hypothesis states that
\begin{equation}
\overline{\rho'\,\uvec{v}'} = - K \unabla \overline{\rho},
\end{equation}
where $K$ is the turbulent diffusivity. Inserting this into the averaged continuity equation yields
\begin{equation}
\frac{\overline{\td} \overline{\rho}}{\overline{\td} t} = - \overline{\rho}\unabla\cdot\overline{\uvec{v}} + \unabla\cdot(K \unabla \overline{\rho}).
\label{e:vdt:cont-w-turb-hyp}
\end{equation}
As no fluctuating quantities remain we will drop the bars in the following. Expressing Eq. \eqref{e:vdt:cont-w-turb-hyp} in terms of SPH operators the following is obtained
\begin{equation}
\frac{\td \rho_a}{\td t} = - \rho_a\textbf{Div}^\gamma_a(\uvec{v}) + \textbf{Lap}^\gamma_a(K,\rho),
\end{equation}
where $\textbf{Lap}^\gamma$ is an SPH discrete operator, here applied to $\rho$ with a diffusion coefficient $K$. If the model by Morris \cite{morris_modeling_1997} is used for the discretisation of the $\textbf{Lap}^\gamma$ operator as in Eq. \eqref{e:sph:lap}, without boundary terms, then the full discretisation reads
\begin{equation}
\frac{\td \rho_a}{\td t} = \rho_a\underset{b\in\mathcal{P}}{\sum}V_b \left(\uvec{v}_{ab} + (K_a+K_b)\rho_{ab}\frac{\uvec{r}_{ab}}{\uvec{r}_{ab}^2}\right)\cdot \unabla_a w_{ab},
\label{e:derived-cont}
\end{equation}
where $\rho_{ab} = \rho_a - \rho_b$. It is common to write the diffusivity term as $K = \nu_T/\sigma_T$, where $\nu_T$ is the turbulent viscosity and $\sigma_T$ is the turbulent Prandtl number. Continuing the analogy with physically-based turbulence, one may use the mixing length model to estimate $\nu_T$ (see \emph{e.g.} Pope \cite{pope_turbulent_2001}), leading to
\begin{equation}
K \sim \frac{1}{\sigma_T}L_m^2\frac{U}{L},
\end{equation}
where $U$ and $L$ are characteristic velocity and length scales, respectively, and $L_m$ is the mixing length. Defining the numerical Mach number as $Ma = U/c_0$, where again $c_0$ is the numerical speed of sound, yields
\begin{equation}
K \sim \frac{Ma}{\sigma_T}L_m^2\frac{c_0}{L}.
\end{equation}
Typically, $\sigma_T \approx 1$, $L_m = L/10$ and in weakly compressible SPH $Ma = 1/10$ \cite{monaghan_smoothed_2005}, which yields
\begin{equation}
K \sim \frac{L}{\Delta r}\frac{c_0\,\Delta r}{10^3}.
\end{equation}
Depending on the resolution, $K$ can thus be given as 
\begin{equation}
K = \frac{c_0\,\Delta r}{\eta},
\label{e:vdt:K-w-alpha}
\end{equation}
where $\eta \approx 10^3 \Delta r/L$ typically has the range of values of $\mathcal{O}(10) - \mathcal{O}(100)$ depending on the ratio of $\Delta r/L$ used to resolve the length scale $L$.\\
Ferrari \emph{et al.} \cite{ferrari_new_2009} proposed a diffusion term which is remarkably similar to the one above (Eqs. \eqref{e:derived-cont} and \eqref{e:vdt:K-w-alpha}). It is based on the theory of Riemann solvers which were first introduced to SPH by Vila \cite{vila_particle_1999}. Ferrari \emph{et al.} use an approximate Riemann solver to obtain the following continuity equation
\begin{equation}
\frac{\td \rho_a}{\td t} = \underset{b\in\mathcal{P}}{\sum}V_b\left(\rho_b \uvec{v}_{ab} + c_{ab}\frac{\uvec{r}_{ab}}{r_{ab}}\rho_{ab}\right)\cdot\unabla_a w_{ab},
\label{e:ferrari-cont}
\end{equation}
where
\begin{equation}
c_{ab} = \max(c_a,c_b),
\label{e:ferrari-upw}
\end{equation}
and
\begin{equation}
c_a = c_0\sqrt{\left(\frac{\rho_a}{\rho_0}\right)^{\zeta-1}},
\label{e:vdt:ca}
\end{equation}
with $\rho_0$ being the reference density and $\zeta$ the exponent of the equation of state \eqref{e:overview:eos}. Comparing Eqs. \eqref{e:derived-cont} and \eqref{e:ferrari-cont} it can be deduced that our model is equivalent to that of Ferrari \emph{et al.} \cite{ferrari_new_2009} if
\begin{equation}
r_{ab}c_{ab} = K_a+K_b.
\label{e:K-d-relation}
\end{equation}
Now a closer examination of $c_a$, given by \eqref{e:vdt:ca}, Eq. \eqref{e:overview:eos} shows that
\begin{equation}
c_a = c_0 \left(\frac{\rho_a}{\rho_0}\right)^{\frac{\zeta-1}{2}} =  \sqrt{\frac{\partial p}{\partial \rho}\bigg|_a},
\end{equation}
so $c_a$ is \deleted{nothing else than} the speed of sound of particle $a$. Thus $K$ has the dimension $m^2/s$ as expected for the turbulent diffusivity term.\\
Comparing Eqs. \eqref{e:K-d-relation} and \eqref{e:vdt:K-w-alpha} shows that the correction as proposed by Ferrari \emph{et al.} \cite{ferrari_new_2009} applies a significantly higher viscosity term than what can be derived from the simple mixing length model. This fact can be of importance when it comes to the use of this correction in the context of turbulent flows, where SPH is used increasingly (\emph{e.g.} Ting \emph{et al.} \cite{ting_simulation_2006}). The turbulent viscosity introduced due to the volume diffusion term by Ferrari \emph{et al.} \cite{ferrari_new_2009} can dissipate more than just the numerical noise and thus have an influence on the energy spectrum of the flow.\\
Finally, note that in the Reynolds averaged context the Navier-Stokes equations would also need to be averaged. This was neglected in the above as the aim was to find a different interpretation for the volume diffusion term by Ferrari \emph{et al.} \cite{ferrari_new_2009} which only acts on the continuity equation.

\subsection{Semi-analytical wall boundary framework}
\label{s:vdt:sa}
After this interpretation the question arises of how this additional numerical diffusion term can be included in the wall boundary formulation as described above. To do so, the flux of the quantity $K \unabla \rho$ has to be investigated in the normal direction of the wall. As the boundary condition on the density implies $\partial \rho/\partial n = 0$ for flows without external forces, this flux is zero as well. Thus, using the Laplacian of Ferrand \cite{ferrand_unified_2012, ferrand_improved_2010} and the divergence given by Eq. \eqref{e:sa:div-sa} the continuity equation with volume diffusion term reads
\begin{equation}
\frac{\td \rho_a}{\td t} = \frac{\rho_a}{\gamma_a}\underset{b\in\mathcal{P}}{\sum}V_b \left(\uvec{v}_{ab} + \frac{c_{ab}}{\rho_a}\rho_{ab}\frac{\uvec{r}_{ab}}{r_{ab}}\right)\cdot \unabla_a w_{ab}- \frac{\rho_a}{\gamma_a}\underset{s\in\mathcal{S}}{\sum}\uvec{v}_{as}\cdot \unabla \gamma_{as}.
\label{e:vdt:mf}
\end{equation}
As can be seen from the above, the volumetric term contains also vertex particles. In turn however, their density is determined from the boundary condition. As this volume diffusion term can be seen as a way of transferring volume from one particle to the next it is important that this term is anti-symmetric, \emph{i.e.} the volume taken from a particle $a$ due to the influence of particle $b$ needs to be added to particle $b$ as a result of the influence of particle $a$. This principle is thus violated if vertex particles are taken into account in the volumetric sum. In 2-D simulations this change has negligible impact on the flow.

\section{Imposing a volume flux in periodic viscous flows}
\label{s:volflux}
\begin{figure}[b!]
\begin{center}
\input{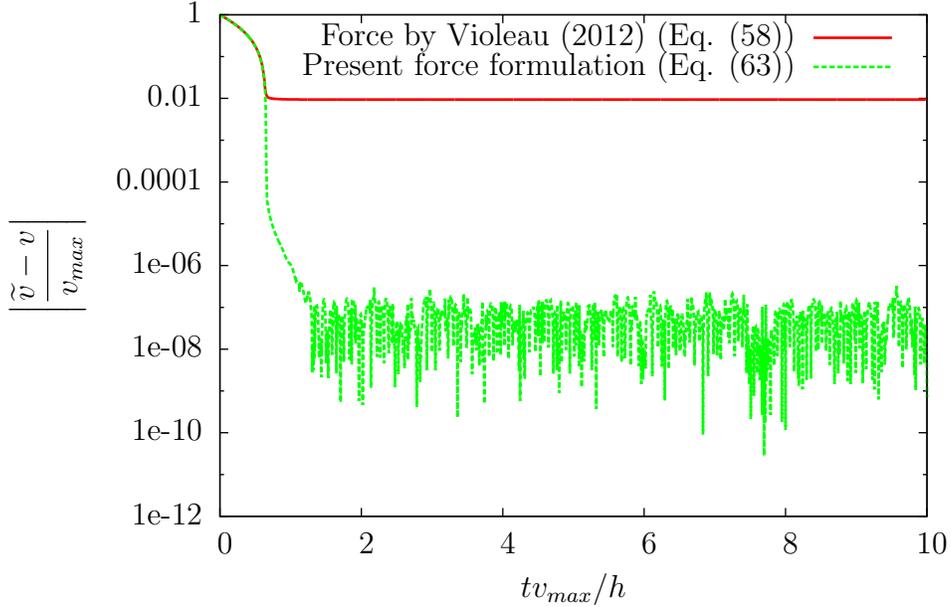}
\end{center}
\caption{Comparison of the error in the bulk velocity of the Poiseuille flow.}
\label{f:volflux:error}
\end{figure}

\added{Related to the issue of solid boundaries are open boundary conditions. In the following the focus will lie on periodic boundaries which itself do not pose an issue with SPH. However, in order to drive such a periodic flow with a fixed volume flux and thus variable force only one formula exists, which shows relatively large errors as will be shown in the following.}
\\
\replaced{Imposing movement in the standard Poiseuille flow is normally achieved with a fixed driving force.}{To impose a movement in the Poiseuille flow a fixed driving force was used.} This \replaced{is}{was} however only possible if the value of the latter was known \emph{a priori}, since the expected velocity profile is known. It is more common to drive a flow with a certain volume flux $Q$. In the following it will be described how to impose a variable driving force based on the expected volume flux.\\
The volume flux $Q$ is defined as
\begin{equation}
Q = \int_A \uvec{v}\cdot \td \uvec{A} = \int_A \uvec{v}\cdot\uvec{n} \td A,
\end{equation}
where $A$ is the cross-section area of the flow, $\uvec{n}$ its normal and $\uvec{v}$ the velocity. To obtain this value in the SPH framework an average over all particles in a slice of the domain will be taken, \emph{i.e.}
\begin{equation}
Q_{SPH} = \frac{1}{\Delta r_A} \underset{b\in\mathcal{P}_A}{\sum} V_b \uvec{v}_b \cdot \uvec{n}_A,
\end{equation}
where $\Delta r_A$ is the width of the slice and the set $\mathcal{P}_A$ contains the particles in the slice. Herein, $\Delta r_A$ is twice the particle spacing $\Delta r$. This can also be written as
\begin{equation}
Q_{SPH} = A \widetilde{v},
\end{equation}
where $\widetilde{v}$ is the cross-averaged velocity.\\
In the book by Violeau \cite{violeau_fluid_2012} this average velocity is used to calculate the force via
\begin{equation}
F^{ext,n} = \frac{v - 2\widetilde{v}^{n-1} + \widetilde{v}^{n-2}}{2\Delta t},
\label{e:volflux:old-force}
\end{equation}
where $n$ is the n-th time-step, $F$ is the force in direction of the normal and $v=Q/A$ the desired velocity. This formula originates from the finite volume community \cite{rollet-miet_simulation_1998}. As can be seen in Figure \ref{f:volflux:error} this value does not converge to the analytical (expected) one. The reason for this deficiency comes from the fact that internal forces are not considered in the above formulation and so the external force is always underestimated. The equilibrium that should be reached is defined by
\begin{equation}
F^{ext} + F^{int} = 0,
\end{equation}
where $F^{ext}$ and $F^{int}$ are the external and internal force respectively (the latter includes pressure and viscous forces). If the system is not in equilibrium the following holds
\begin{equation}
F^{ext,n} + F^{int,n} = \frac{\widetilde{v}^{n} - \widetilde{v}^{n-1}}{\Delta t^n}.
\label{e:volflux:acc}
\end{equation}
Ideally the velocity reached at time $n$ is equal to the desired velocity $v = Q/A$. So rewriting the above yields
\begin{equation}
F^{ext,n} = \frac{v - \widetilde{v}^{n-1}}{\Delta t^n} - F^{int,n}.
\end{equation}
Clearly $F^{int,n}$ is not available but it can be assumed to vary only a little between two consecutive time-steps, \emph{i.e.} $F^{int,n} \approx F^{int,n-1}$. The latter value can then be calculated using Eq. \eqref{e:volflux:acc} to give
\begin{equation}
F^{ext,n} \approx \frac{v - \widetilde{v}^{n-1}}{\Delta t^n} - \frac{\widetilde{v}^{n-1} - \widetilde{v}^{n-2}}{\Delta t^{n-1}} + F^{ext,n-1}.
\end{equation}
Finally rearranging the above yields the following new formula for calculating the external force:
\begin{equation}
F^{ext,n} = \frac{v - 2\widetilde{v}^{n-1} + \widetilde{v}^{n-2}}{\Delta t} + F^{ext,n-1},
\label{e:volflux:new-force}
\end{equation}
where it is assumed that the time-step $\Delta t$ is constant. The first term in the right-hand side of Eq. \eqref{e:volflux:new-force} is twice that of Eq. \eqref{e:volflux:old-force}. Thus \eqref{e:volflux:old-force} amounts to considering that $F^{ext}$ is independent of time, which is a crude approximation.\\
In Figure \ref{f:volflux:error} the two different means of imposing a driving force based on a volume flux are compared in the Poiseuille flow case. To do so the relative error in the bulk velocity is plotted over time. In Figure \ref{f:volflux:error} it can be seen that with this new formulation the external force converges much more closely to the theoretical value. In the case of the Poiseuille flow it is not necessary to impose a volume flux but instead the analytical force can be used. However, in cases where the analytical value of the internal (viscous) force is not known \emph{a priori} the above formulation provides the means to impose a volume flux which converges to the desired value.\\
As it can be seen \added{from \figurename~\ref{f:volflux:error}} the present approach reduced the error by about five orders of magnitude. The error obtained with the original formulation is close to 1\% which is not negligible.

\section{Generalization of wall boundary conditions}
\label{s:bcs}
\added{After this analysis of the unified semi-analytical wall boundary conditions and the reinterpretation of the volume diffusion term the focus in the following two sections will shift towards novel developments which expand the boundary model.}

\subsection{Theory}
As mentioned in Section \ref{s:overview}, to satisfy von Neumann boundary conditions Ferrand \emph{et al.} \cite{ferrand_unified_2012} showed a first-order approximation approach (see Eqs. \eqref{e:overview:neumann-bcs} and \eqref{e:overview:alpha-v}). In the following this approach shall be generalized to arbitrary orders of accuracy and to Robin boundary conditions, which include Neumann boundary conditions as particular case. Ryan \emph{et al.} \cite{ryan_novel_2010} implemented Robin boundary conditions for SPH with the use of an additional source term in the governing equation. In this section a different approach will be shown that enables Robin boundary conditions to be imposed directly.\\
Such a boundary condition is given by
\begin{equation}
\left(\mu_1 f + \mu_2 \frac{\partial f}{\partial n}\right)\bigg|_{\partial\Omega} = \mu_3,
\end{equation}
for an arbitrary scalar field $f$ with given $\mu_1,\mu_2\neq0,\mu_3$. To impose Neumann boundary conditions $(\mu_1,\mu_2)$ take the values $(0,1)$. The above can be rewritten as
\begin{equation}
\frac{\partial f}{\partial n} = \frac{\mu_3-\mu_1f}{\mu_2}.
\end{equation}
\begin{figure}[htb]
\begin{center}
\includegraphics[width=0.5\textwidth]{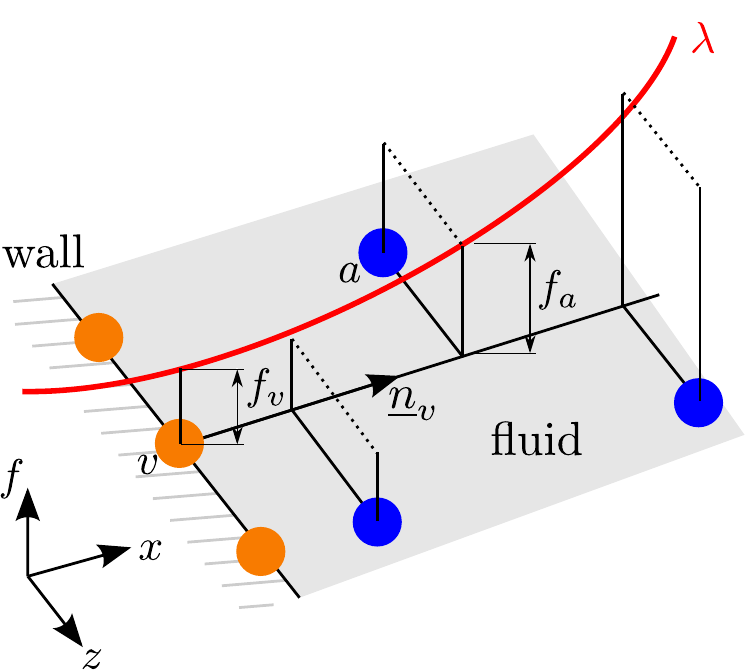}
\end{center}
\caption{Illustration of the arbitrary order Robin boundary conditions for SPH. Blue particles denote fluid particles and particle $v$ in orange is a vertex particle under consideration.}
\label{f:aorbcs:ill}
\end{figure}
The key idea is to use a weighted linear least squares approximation of the desired field in the direction of the normal $\uvec{n}_v$ of a vertex particle $v$, which is defined as the average over the two adjacent segment normals. This implies that the problem is considered locally as one dimensional projected onto the normal $\uvec{n}_v$ of the vertex particle $v$. The procedure presented in detail in the following is illustrated in Fig. \ref{f:aorbcs:ill}.\\
Let $m$ be the desired order of approximation and define a polynomial $\lambda$ as
\begin{equation}
	\lambda(x) = \underset{i=2}{\overset{m}{\sum}}\beta_i x^i + \frac{\mu_3-\mu_1\beta_1}{\mu_2} x + \beta_1,
	\label{e:bcs:lambda}
\end{equation}
$x$ being the normal distance to the wall.\\
We set $\beta_1 = f\big|_{\partial\Omega}$, \emph{i.e.} $\beta_1$ is the value of $f$ prescribed at the wall. The polynomial can also be written as
\begin{equation}
	\lambda(x) = \underset{i=2}{\overset{m}{\sum}}\beta_i x^i + \frac{\mu_3}{\mu_2} x + \beta_1\left(1-\frac{\mu_1}{\mu_2}x\right).
\end{equation}
This represents an approximation of $f$ in the direction of the normal $\uvec{n}_v$ with $f_v = \lambda(0)$. For a fluid particle $a \in \mathcal{F}$ and $x_a = \uvec{r}_{av}\cdot\uvec{n}_v$ the value of $\lambda_a$ is thus given as $\lambda(\uvec{r}_{av}\cdot\nolinebreak\uvec{n}_v)$ when no external forces are present. Let $\umat{X} \in \mathbb{M}(|\mathcal{F}|,m)$ be a matrix defined as
\begin{equation}
\umat{X}_{aj} = \left\{ \begin{array}{rl}
1-\frac{\mu_1}{\mu_2} (\uvec{r}_{av}\cdot\uvec{n}_v) &\mbox{ if $j = 1$,} \\
(\uvec{r}_{av}\cdot\uvec{n}_v)^j &\mbox{ if $j > 1$,}
                \end{array} \right.
\end{equation}
\added{where $j$ is the second index of the matrix $\umat{X}$,} $|\mathcal{F}|$ is the number of fluid particles and defining $\uvec{y} \in \mathbb{R}^{|\mathcal{F}|}$ as
\begin{equation}
y_a = f_{a}-\frac{\mu_3}{\mu_2} (\uvec{r}_{av}\cdot\uvec{n}_v).
\label{e:aorbcs:def-y}
\end{equation}
The goal is to minimize $f_a - \lambda_a$. Hence, we are looking for a solution of the over constrained system of linear equations
\begin{equation}
\umat{X}\cdot\uvec{\beta} = \uvec{y},
\end{equation}
where $\uvec{\beta}$ is the vector of components $\beta_i$ and the dot represents a single contraction, in this case the matrix multiplication. A common approach to this type of over-constrained problem is to use the least squares method which can be solved via
\begin{equation}
\umat{X}^T\cdot\umat{X}\cdot\uvec{\beta} = \umat{X}^T\cdot\uvec{y}.
\end{equation}
In order to put more weight on particles closer to the vertex particle, the SPH kernel is used to construct a weighted least squares interpolation. To include this into the above let \added{$\delta_{ab}$ be the Kronecker delta and }$\umat{\Lambda} \in \nolinebreak \mathbb{M}(|\mathcal{F}|)$, a square matrix, be defined with elements given by
\begin{equation}
\umat{\Lambda}_{ab} = \delta_{ab}V_{a}w_{va},
\end{equation}
(there is no summation over indices here). Then the weighted least squares interpolation can be found by solving
\begin{equation}
\umat{X}^T\cdot\umat{\Lambda}\cdot\umat{X}\cdot\uvec{\beta} = \umat{X}^T\cdot\umat{\Lambda}\cdot\uvec{y}.
\end{equation}
Due to the fact that the problem is one dimensional the matrix on the left-hand side is of size $m\times m$ and can be inverted easily. Of interest is $\lambda(0)$ which is $\beta_1$. Thus finally
\begin{equation}
f_v = \lambda(0) = \beta_1 = ((\umat{X}^T\cdot\umat{\Lambda}\cdot\umat{X})^{-1}\cdot\umat{X}^T\cdot\umat{\Lambda}\cdot\uvec{y})_1,
\label{e:aorbcs:fv}
\end{equation}
where the subscript $1$ on the right hand side refers to the first component of the vector in brackets.\\
If $m=1$ the above reduces to the equation shown in Eq. \eqref{e:overview:neumann-bcs} thus showing that this indeed is a generalization to arbitrary order. Note that the matrix that needs to be inverted $(\umat{X}^T\cdot\umat{\Lambda}\cdot\umat{X})$ is contained in $\mathbb{M}(m)$ and thus to obtain second-order boundary conditions the inversion of a $2 \times 2$ matrix is required in two dimensions.\\
Finally, two things should be noted. Firstly, the matrix is non-degenerate if at least $m$ fluid particles are in the neighbourhood of $v$ and that they have different values of $(\uvec{r}_{av}\cdot\uvec{n}_v)$. Secondly, the formulation as presented above is independent of the method used to describe the walls, so it could also be used \emph{e.g.} for the SPH ghost particle approach \cite{colagrossi_numerical_2003}.
\\
\added{In comparison to the present method where the boundary condition is directly imposed, Ryan \emph{et al.} \cite{ryan_novel_2010} add an extra term to the governing equations. This work is an extension of Ferrand \emph{et al.} \cite{ferrand_unified_2012} such that it is now possible to specify Robin boundary conditions and not only Neumann ones. Moreover, these can be imposed to arbitrary order.}
\subsection{The wave equation as an example}
\label{s:aorbcs:wave}
\begin{figure}[htb]
\begin{center}
\subfigure[$t=0.1$]{
\includegraphics[width=0.46\textwidth]{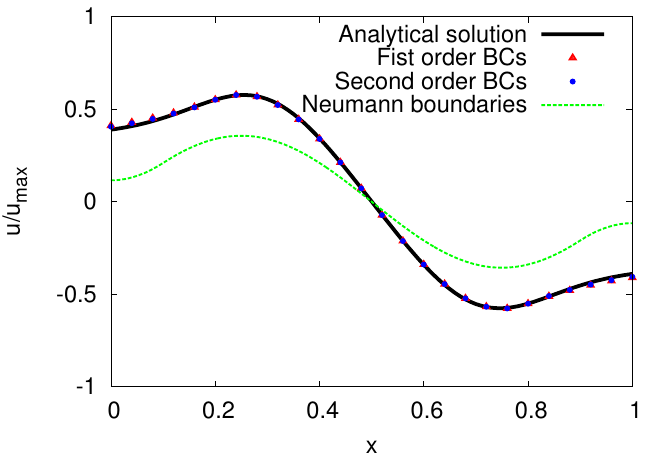}
\label{f:aorbcs:wave:bad1}
}
\subfigure[$t=0.22$]{
\includegraphics[width=0.46\textwidth]{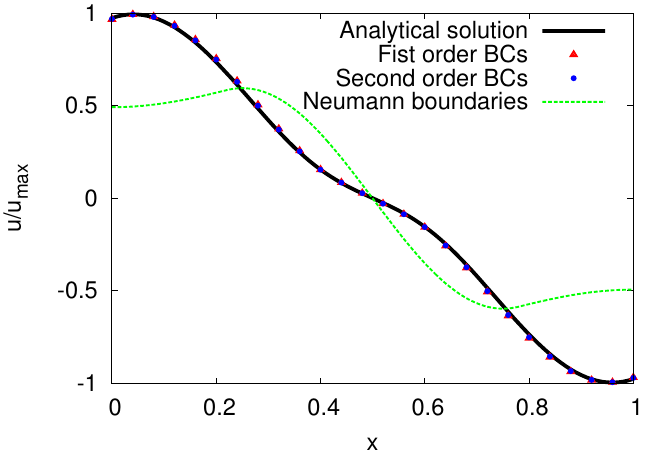}
\label{f:aorbcs:wave:good1}
}
\\
\subfigure[$t=0.7$]{
\includegraphics[width=0.46\textwidth]{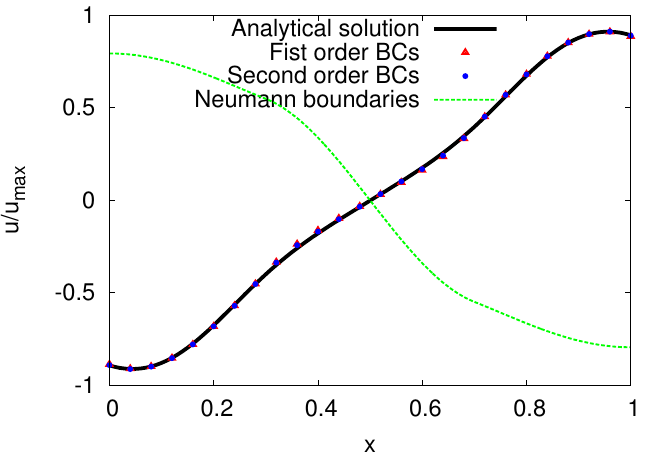}
\label{f:aorbcs:wave:good2}
}
\subfigure[$t=0.81$]{
\includegraphics[width=0.46\textwidth]{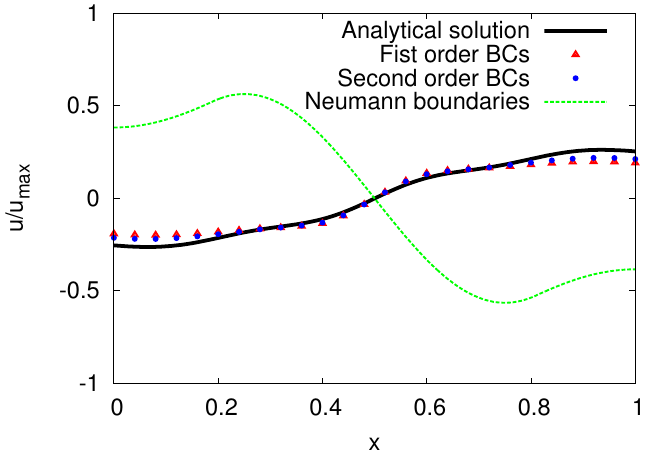}
\label{f:aorbcs:wave:bad2}
}
\end{center}
\caption{Wave equation \eqref{e:aorbcs:waveeq} with Robin boundary conditions \eqref{e:aorbcs:wavebcs} at four different time-steps. \added{Analytical solution given by Eq. \eqref{e:aorbcs:sol_ana}. For the SPH solution, the matrix size in Eq. \eqref{e:aorbcs:fv} is given by first ($m=1$) or second ($m=2$) order according to Eq. \eqref{e:bcs:lambda}}}
\label{f:aorbcs:wave:time}
\end{figure}
In this section it will be shown that the above formulation indeed works, i.e. that Robin boundary conditions can be enforced. For this purpose, a wave equation is considered in 1 dimension:
\begin{equation}
\frac{\partial^2 u}{\partial t^2} = \frac{\partial^2 u}{\partial x^2},
\label{e:aorbcs:waveeq}
\end{equation}
with boundary conditions
\begin{equation}
\frac{\partial u}{\partial n}(\{0,1\},t) = u(\{0,1\},t),
\label{e:aorbcs:wavebcs}
\end{equation}
and initial conditions
\begin{equation}
u(x,0) = 0\qquad\mbox{and}\qquad\frac{\partial u}{\partial t}(x,0) = \sin(2\pi x).
\end{equation}
The domain $[0,1]$ is discretized with an initial particle distance $\Delta r = 0.01$ and the particles are fixed. The second-order spatial differential operator is again discretized by using the Laplacian as given in Eq. \eqref{e:sph:lap}. The temporal derivative is discretized by a second-order finite difference scheme. The analytical solution \cite{haberman_applied_2004} of this problem as a function of the dimensionless variables $x$ and $t$ is given by
\begin{equation}
u_{ana}(x,t) = \underset{i=1}{\overset{\infty}{\sum}} \alpha_i \sin(\kappa_i t) \left[\sin(\kappa_i x) + \kappa_i \cos(\kappa_i x)\right],
\label{e:aorbcs:sol_ana}
\end{equation}
where $\kappa_i$ is given as the i-th root of
\begin{equation}
\tan(\kappa_i) = \frac{2\kappa_i}{\kappa_i^2-1},
\end{equation}
and the $\alpha_i$ are uniquely determined by the initial condition.
\\
First, a qualitative view on the solution is given in Fig. \ref{f:aorbcs:wave:time} at times $t=\{0.1,\,0.22,\,0.7,\,0.82\}$. It can be seen that the main features of the field $u$ are reproduced by the SPH solution and that, due to the  new boundary condition formulation, the Robin boundary conditions are correctly imposed as we observed a distinct difference in the solution when using Neumann boundary conditions, \emph{i.e.} $\mu_1=0$. In order to illustrate this the analytical solution for the Neumann boundary case was superimposed in Fig. \ref{f:aorbcs:wave:time}. The snapshots in Figs. \ref{f:aorbcs:wave:bad1} and \ref{f:aorbcs:wave:bad2} are plotted \deleted{during} at instants where the error on the boundary is maximal. Compared to these snapshots, the error between the SPH solution and the analytical solution is smaller throughout the domain at intermediate times (Figs. \ref{f:aorbcs:wave:good1} and \ref{f:aorbcs:wave:good2}).
\begin{figure}[htb]
\begin{center}
\includegraphics[width=0.6\textwidth]{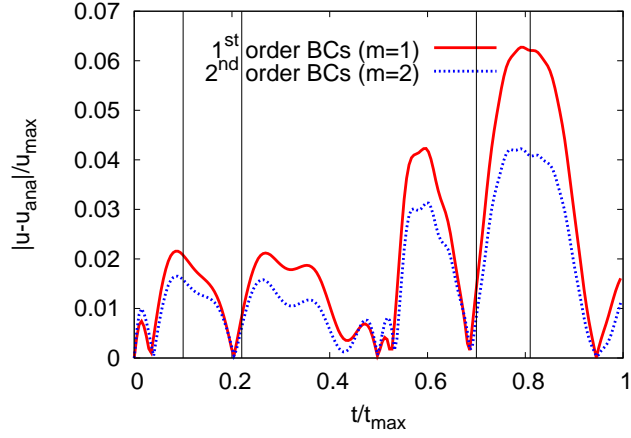}
\end{center}
\caption{Wave equation with Robin boundary conditions: Relative error of the SPH solution over time at the boundary.}
\label{f:aorbcs:wave:error}
\end{figure}
\\
There is a notable difference in the SPH solution depending on whether the boundary conditions are of first or second order, i.e. whether $m=1$ or $m=2$ in Eq. \eqref{e:bcs:lambda}. To quantify this more precisely consider Fig. \ref{f:aorbcs:wave:error} where the error at the boundary is plotted over time. It can be observed that the error is sensitive to the order of the boundary condition and that it can be reduced by up to 30\% by choosing a second-order approximation. The vertical bars in Fig. \ref{f:aorbcs:wave:error} show the instants that were shown in Fig. \ref{f:aorbcs:wave:time}.

\section{Still water with a free-surface}
\label{s:sw}
In the following, two of the above presented improvements will be reviewed in the context of still water free-surface \replaced{evolution}{flows}. The first section below will focus on the volume diffusion term and the second one on the arbitrary order boundary conditions.

\subsection{Modification of the volume diffusion term}
\label{s:sw:vdt}
Using the volume diffusion term presented in Section \ref{s:vdt} in a simulation of still water shows that the term as shown above will not have a zero contribution towards the density equation. This is caused by the fact that the boundary terms in Eq. \eqref{e:vdt:mf} do not vanish when gravity is present. However, there are no segments on the free-surface and the problem can thus not be resolved by adding a boundary term.
\begin{figure}[htb]
\begin{center}
\subfigure[Free-surface position over time]{
  \scalebox{0.55}{ \input{fs_position} }
}
\subfigure[Flow without and with free-surface correction]{
  \includegraphics[width=0.49\textwidth]{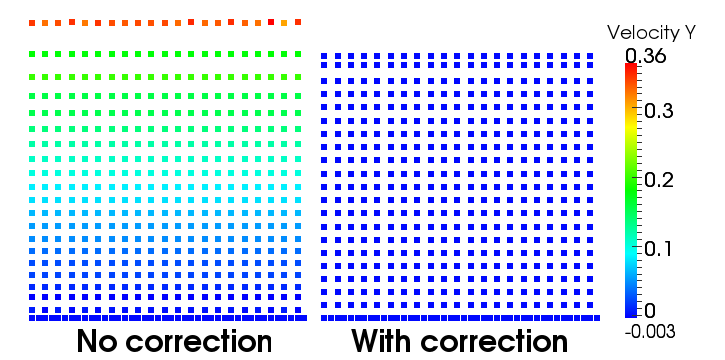}
}
\end{center}
\caption{Position of free-surface in infinite open channel at rest.}
\label{f:sw:vdt:fs-pos}
\end{figure}
\\
In order to compensate for this deficiency the correction can be modified so that it reads
\begin{equation}
\frac{\td \rho_a}{\td t} = \frac{\rho_a}{\gamma_a}\underset{b\in\mathcal{P}}{\sum}V_b\left(\uvec{v}_{ab} + \frac{c_{ab}}{\rho_a}\varrho_{ab}\frac{\uvec{r}_{ab}}{r_{ab}}\right)\cdot \unabla_a w_{ab} - \frac{\rho_a}{\gamma_a}\underset{s\in\mathcal{S}}{\sum}\uvec{v}_{as}\cdot\unabla\gamma_{as},
\label{e:vdt:fscorr}
\end{equation}
in place of Eq. \eqref{e:vdt:mf} where \added{the density difference $\rho_{ab}$ was replaced by the modified density difference} $\varrho_{ab}$ \added{which} is \deleted{a modified density difference} given by
\begin{equation}
\varrho_{ab} = \rho_a - \rho_b - \frac{g \rho_0}{c_0^2}(z_b-z_a).
\label{e:vdt:modrho}
\end{equation}
where $g$ is the gravitational constant and $z$ the vertical elevation. This is a linear approximation of the difference between the non-hydrostatic densities of two particles. The external force in the formula above is gravity, but adaptation to other forces is straightforward.
\\
In Section \ref{s:vdt:sa} it was remarked that the volume diffusion term should be anti-symmetric with respect to particles $a$ and $b$ in order to avoid that the global volume is changed. Clearly Eq. \eqref{e:vdt:modrho} obeys this principle. A similar correction was simultaneously proposed by Antuono \emph{et al.} \cite{antuono_numerical_2012}.
\\
To illustrate the difference between using the traditional $\rho_{ab} = \rho_a - \rho_b$ and $\varrho_{ab}$ consider Fig. \ref{f:sw:vdt:fs-pos}, which shows the non-dimensionalized position of the free-surface of an open channel-flow at rest where the initial density is set to $\rho_0$. In the plot the time was renormalized by $h_{swl}/c_0$, where $h_{swl} = 1$ is the still water-level and $c_0 = 10$ the numerical speed of sound. It can be seen that without the above modification the free-surface detaches due to a transfer of volume from the denser lower part to the upper part of the fluid. The modification proposed above clearly avoids this issue keeping the free-surface elevation almost constant as expected. The decrease is due to the initial condition and the weak compressibility of the fluid.

\subsection{Modification of the boundary interpolation}
\label{s:sw:aorbcs}
In this section the generalized boundary conditions presented in Section \ref{s:bcs} will be investigated in the presence of gravity. This means that the function $f$ will be equal to the pressure for which the classical Neumann boundary condition ($\partial p/\partial n = \rho \uvec{g}\cdot\uvec{n}$) will be applied.
\begin{figure}[htb]
\begin{center}
\includegraphics[width=0.7\textwidth]{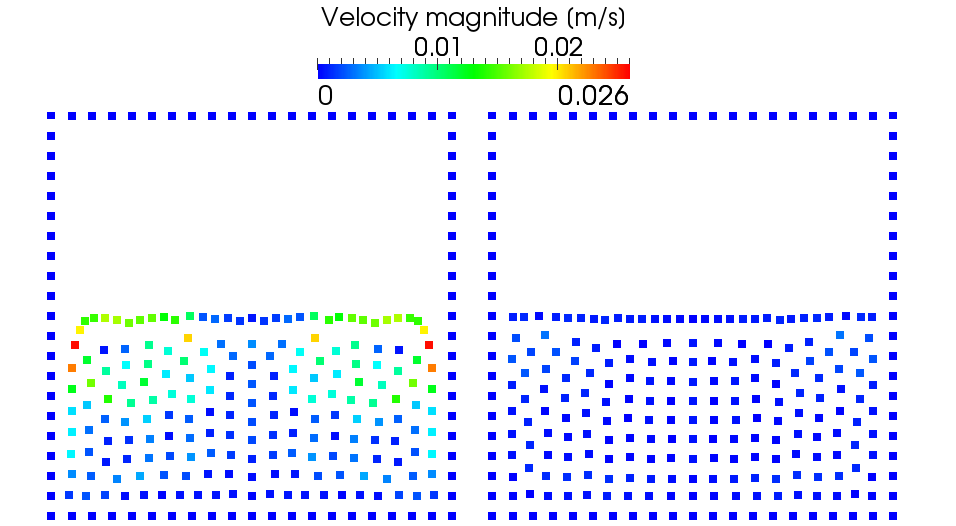}
\end{center}
\caption{Still water in a closed tank (left: without correction \eqref{e:sw:aorbcs:nocor}, right: with correction \eqref{e:sw:aorbcs:extfcor}).}
\label{f:sw:aorbcs:tank}
\end{figure}
\\
The approach presented in Section \ref{s:bcs} will produce unsatisfactory results at the intersection of free-surface and a wall as tangential variations are not neglected. The constraint above (Eq. \eqref{e:aorbcs:def-y}) reads
\begin{equation}
\lambda_a \approx y_a = p_a.
\label{e:sw:aorbcs:nocor}
\end{equation}
In order to neglect tangential variations for external forces such as gravity, the proper constraint is given by
\begin{equation}
\lambda_a \approx y_a = p_a - \rho_a\left[\uvec{r}_{av}-(\uvec{r}_{av}\cdot\uvec{n}_v)\uvec{n}_v\right]\cdot\uvec{g}.
\label{e:sw:aorbcs:extfcor}
\end{equation}
In Fig. \ref{f:sw:aorbcs:tank} the difference between Eqs. \eqref{e:sw:aorbcs:nocor} and \eqref{e:sw:aorbcs:extfcor} is shown. The resolution is chosen to be relatively low in order to highlight the impact of the proposed correction. The picture on the left hand side shows particle movement which is an order of magnitude larger than the one on the right hand side which demonstrates the corrected approximation. Similar to the velocity field, the pressure prediction is improved as well by lowering the magnitude of pressure waves originating from this corner. To explain the formula presented above consider the setup in Fig. \ref{f:sw:aorbcs:tank} with a perfect hydrostatic pressure distribution. Now we look at a vertex particle on a vertical wall which is located next to the free-surface. When constructing the polynomial $\lambda$ (Eq. \eqref{e:bcs:lambda}) as given in Section \ref{s:bcs} the fluid particles considered for the approximation all have a pressure greater or equal to zero. This causes the pressure of the vertex particle to be greater than zero, although its theoretical value is zero. This in turn causes a repulsive force that can be seen on the left hand side of Fig. \ref{f:sw:aorbcs:tank}. If, on the other hand, the hydrostatic part is subtracted from the fluid particles as in Eq. \eqref{e:sw:aorbcs:extfcor} then all fluid particles used for the approximation of $\lambda$ will have zero pressure and thus the vertex particle has the correct pressure.

\section{Dam-break with wedge}
\label{s:fs}
\begin{figure}[htb]
\centering
\input{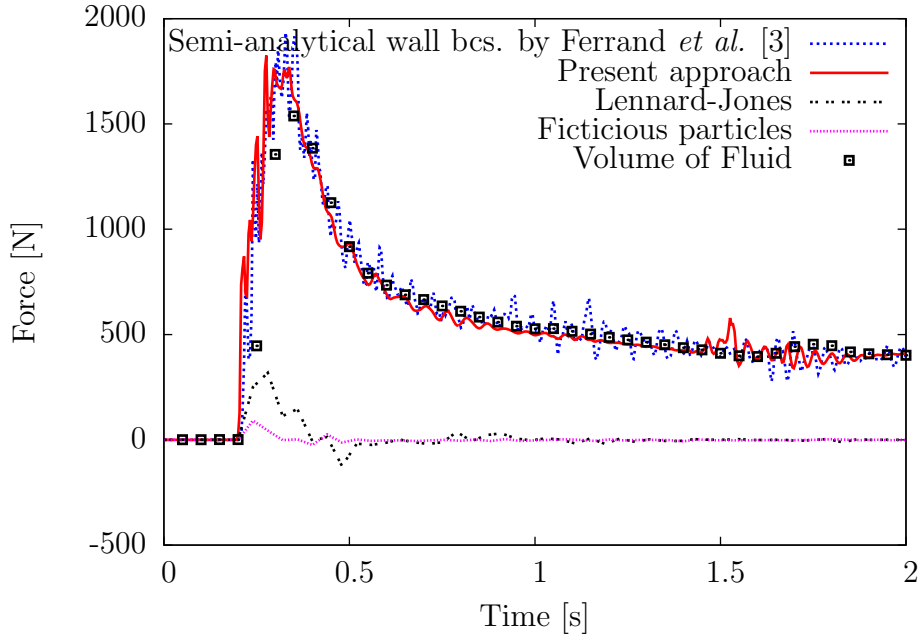}
\caption{Schematic dam-break on a wedge: Comparison of forces on left wedge wall.}
\label{f:2d:dbw-force}
\end{figure}
\begin{figure}[p]
\centering
\subfigure[VOF, t=0.5 s]{
\includegraphics[width=0.46\textwidth]{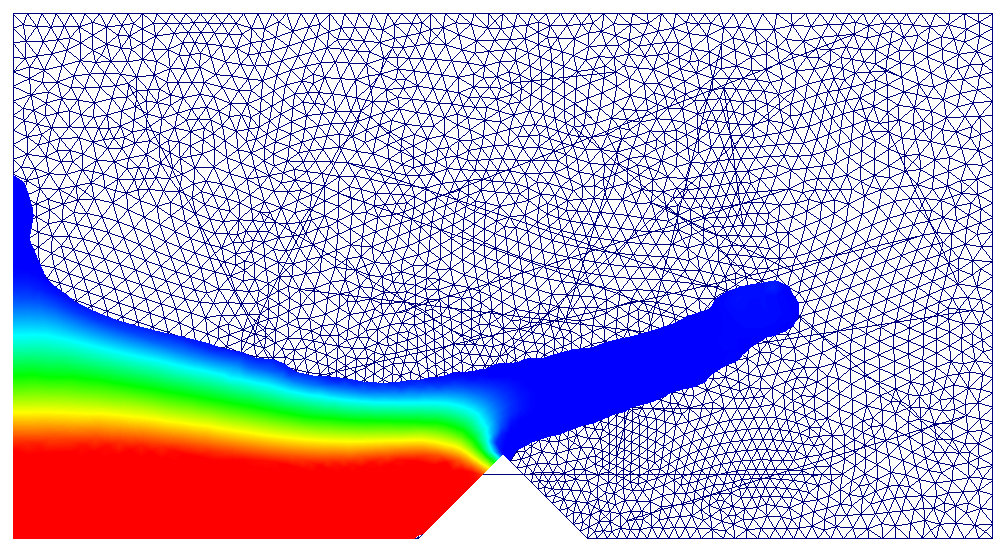}
}
\subfigure[SPH, t=0.5 s]{
\includegraphics[width=0.46\textwidth]{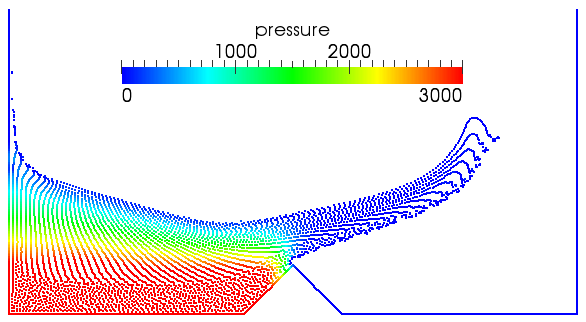}
}\\
\subfigure[VOF, t=0.75 s]{
\includegraphics[width=0.46\textwidth]{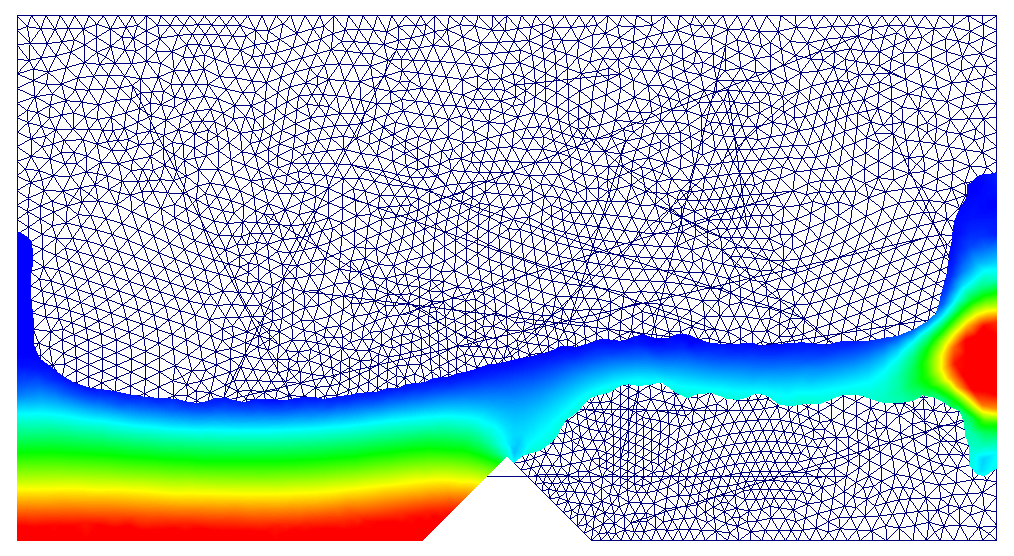}
}
\subfigure[SPH, t=0.75 s]{
\includegraphics[width=0.46\textwidth]{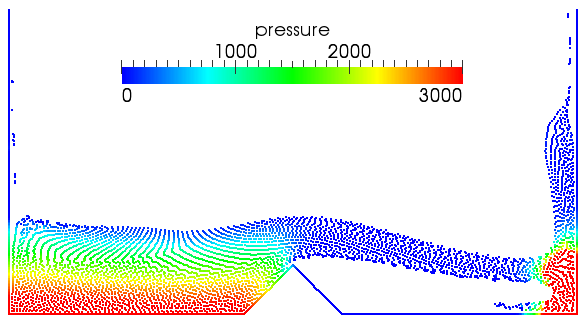}
}
\\
\subfigure[VOF, t=1.1 s]{
\includegraphics[width=0.46\textwidth]{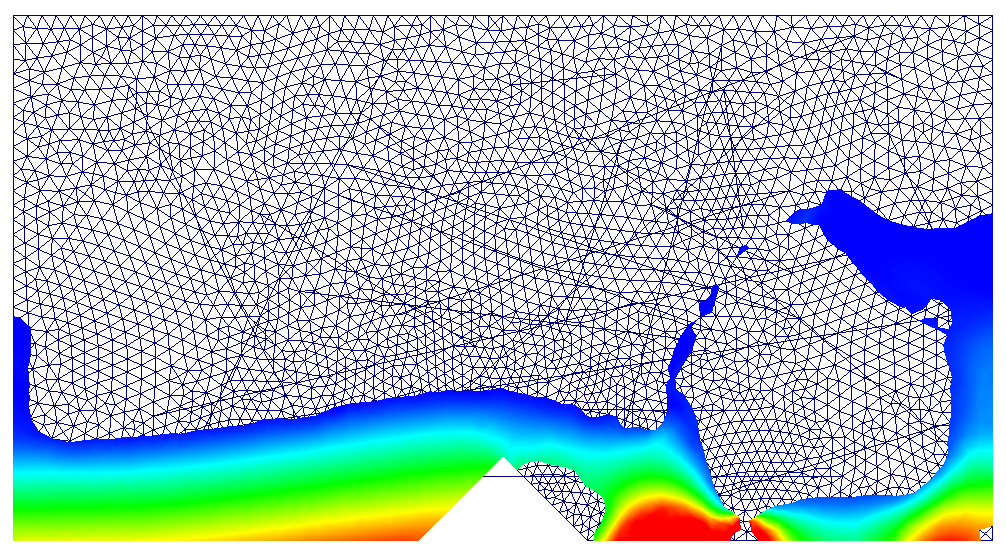}
}
\subfigure[SPH, t=1.1 s]{
\includegraphics[width=0.46\textwidth]{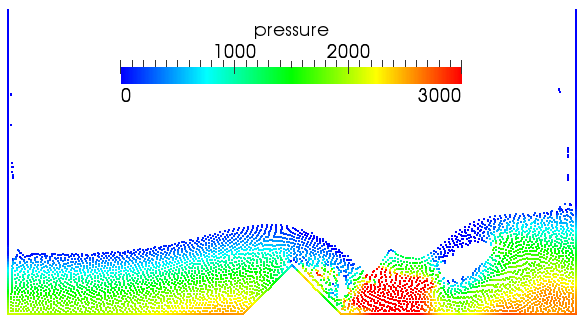}
}\\
\subfigure[VOF, t=2.5 s]{
\includegraphics[width=0.46\textwidth]{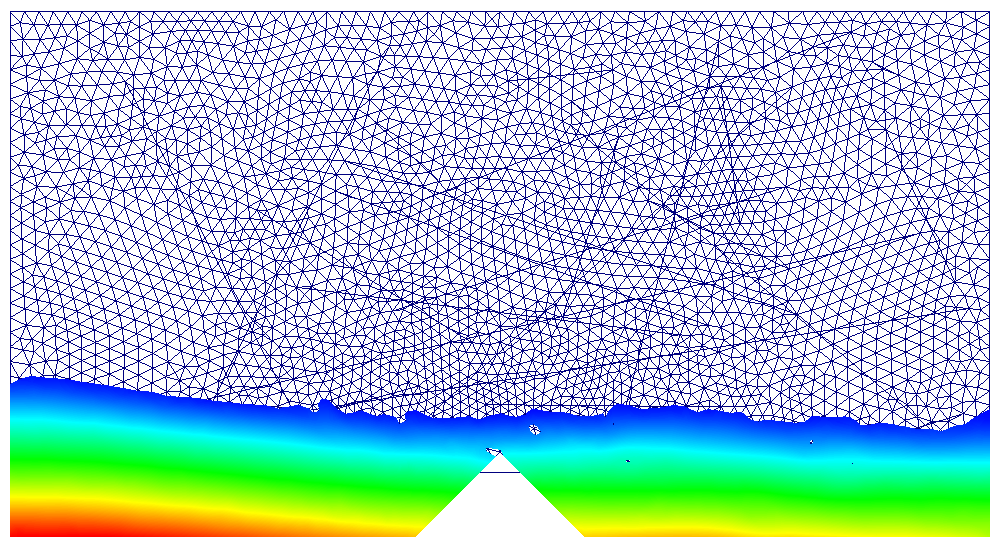}
}
\subfigure[SPH, t=2.5 s]{
\includegraphics[width=0.46\textwidth]{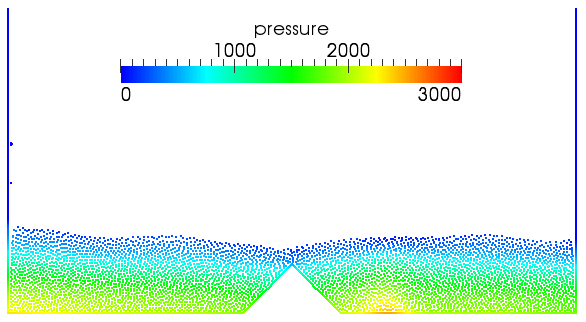}
}
\caption{Comparison between VOF and SPH.}
\label{f:2d:dbw}
\end{figure}
\noindent \added[remark=Reviewer 1: Remark 18]{After having analyzed and extended the present wall boundary conditions a final simulation shall be performed. This uses most of the theoretical results presented above in a more complex free-surface flow.}
\\
\replaced[remark=Reviewer 1: Remark 18]{A}{To study a more complex free-surface flow, a} schematic dam-break over a wedge will be simulated with the same geometry as used in \cite{ferrand_unified_2012}. The initial volume of water is 1 m high and 0.5 m wide and the dynamic viscosity is set to $\nu = 0.01 m^2/s$ resulting in a Reynolds number of approximately 140 based on the maximum velocity. The box has a length of 2.2 m where the wedge begins after 0.85 m with a side-length of 0.25 m. The force will be calculated as the integral of the pressure along the left wedge wall. A Volume-of-Fluid (VOF) simulation is taken as reference solution (OpenFoam \cite{_openfoam_2012}). It should be noted that the latter is a multiphase simulation and thus some discrepancies are to be expected when compared to the SPH single-phase simulation as illustrated in Fig. \ref{f:2d:dbw}. As shown, the traditional boundary conditions using fictitious particles \cite{dalrymple_sph_2001} or the \replaced[remark=Reviewer 2: Remark 9]{Lennard}{Leonard}-Jones potential \cite{monaghan_simulating_1994} fail in predicting the force. Comparing the approach by Ferrand \textit{et al.} \cite{ferrand_unified_2012} with the present one, it can be seen that the volume diffusion term successfully reduces the numerical noise, while still showing closer agreement with VOF in Fig. \ref{f:2d:dbw-force}.
\begin{figure}[t!]
\centering
\subfigure[]{
\includegraphics[width=0.6\textwidth]{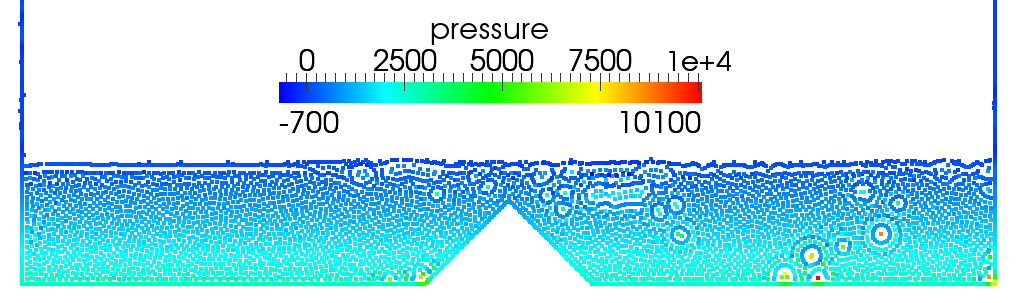}
\label{f:fs:wo-fscor}
}
\\
\subfigure[]{
\includegraphics[width=0.6\textwidth]{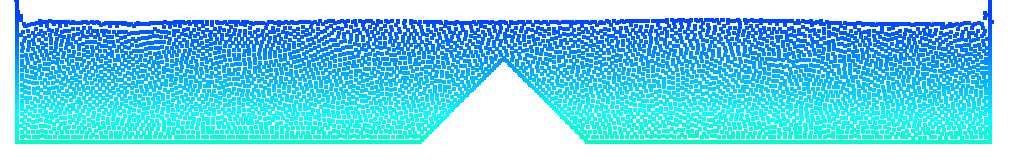}
\label{f:fs:w-fscor}
}
\caption{Steady flow on a wedge: Without (Fig. \ref{f:fs:wo-fscor}) and with (Fig. \ref{f:fs:w-fscor}) the volume diffusion term \eqref{e:fs:ticont-w-vdt} to avoid particle repulsion.}
\label{f:fs:fscor-ti-cont-eq}
\end{figure}\\
Finally, Fig. \ref{f:fs:wo-fscor} shows the steady state solution of the dam break with the time-independent continuity equation (Eq. \eqref{e:overview:ti-cont-eq}) without the heuristic free-surface correction. Particles that were initially on the free surface have retained their larger volumes producing the strange bubbles and unphysical pressures. Following the developments in Section \ref{s:vdt:idea} the integrated-in-time volume diffusion term was added, so that the continuity equation now reads
\begin{equation}
\gamma_a^n \rho_a^n = \gamma_a^{n-1} \rho_a^{n-1} + \underset{b\in\mathcal{P}}{\sum}m_b (w_{ab}^n-w_{ab}^{n-1}) + \Delta t\underset{b\in\mathcal{F}}{\sum}m_b c_{ab}\frac{\uvec{r}_{ab}}{r_{ab}}\frac{\varrho_{ab}}{\rho_b}\unabla_a w_{ab},
\label{e:fs:ticont-w-vdt}
\end{equation}
where the superscripts refer to the time iteration. Note that again the gravitationally corrected volume diffusion term (Eq. \eqref{e:vdt:modrho}) is used through $\varrho_{ab}$. The continuity equation above is equivalent to using $\textbf{Div}^{\gamma,k}$ (Eq. \eqref{e:sa:div-sa}) with $k=1$ so for consistency $\textbf{Grad}^{\gamma,1}$ (Eq. \eqref{e:sa:grad-sa}) has to be used for the discretisation of the pressure gradient.\\
From Fig. \ref{f:fs:w-fscor} it can thus be concluded that the heuristic free-surface correction is no longer required (contrary to \cite{ferrand_unified_2012}) as the volume diffusion term successfully redistributes the higher volume of particles initially on the free surface once they become entrained in the fluid body.

\section{Conclusion}
\added{This paper has investigated SPH boundary conditions for wall-bounded, potentially turbulent, flows within a semi-analytical framework. Three distinct but equally important areas have been investigated.
\\
The semi-analytical wall boundary conditions for SPH introduced by Ferrand \emph{et al.} \cite{ferrand_unified_2012} have been further developed where the skew-adjoint property of discrete operators was examined both theoretically and numerically. It was shown that the skew-adjoint property does not hold in the discrete case leading to errors in the conservation of energy and demonstrated for 2-D channel flows. As shown by Morinishi \emph{et al.} \cite{morinishi_fully_1998} for non-uniform grids, exact conservation is not required if errors are small and remain bounded. A detailed error analysis of the conservation of different boundary conditions would be of interest for further research.
\\
Another general issue with SPH is instability within the method that manifests itself as numerical noise. As shown in this paper the noise can be explained by analogy with a Reynolds-averaged continuity equation which is shown to be equivalent to the density diffusion introduced by Ferrari \emph{et al.} \cite{ferrari_new_2009} which used an approximate Riemann solver. This interpretation justifies the addition of a constant that depends on the relative resolution. As the volume diffusion term introduces artificial viscosity this constant prevents excessive damping which would be problematic such as in the simulation of turbulent cases.
\\
With the aim of simulating turbulent flows in periodic geometries, a novel formulation was presented to prescribe a variable driving force with an imposed volume flux which improves the predicted flow rate by 5 orders of magnitude.  Additionally, the Neumann boundary conditions by Ferrand \emph{et al.} \cite{ferrand_unified_2012} were generalized to arbitrary orders of interpolation and Robin-type boundary conditions. The formulation was shown to impose Robin boundary conditions correctly thereby extending their potential application. Finally, two modifications to the boundary conditions as well as the volume diffusion term were presented in order to correctly deal with free-surface flows, reducing unphysical velocities at the surface by at least an order of magnitude for still water.
\\
The new numerical scheme was demonstrated for a dam-break flow over a wedge showing the capabilities of the present improved model compared to a well-known VOF code. This simulation was also used to highlight the fact that the volume diffusion term can correct the free surface when using the time-independent continuity equation, as proposed by Ferrand \emph{et al.} \cite{ferrand_unified_2012}}

\section*{Acknowledgements}
The authors would like to thank Dominique Laurence for the constructive conversations and Christophe Kassiotis for providing the Volume of Fluids simulations. Thanks also goes to Thomas Klose who provided the analytical solution for the wave equation with Robin boundary conditions.


\bibliographystyle{model1-num-names}
\bibliography{2d-main.bib}

\appendix
\section{Derivation of skew-adjointness}
\label{s:sa-app}
In the following the left hand side of Eq. \eqref{e:sa:sa-ana} will be analyzed when replacing the nabla operator with the gradient and divergence given by Eqs. \eqref{e:sa:gradcont} and \eqref{e:sa:divcont}. Initially the part containing the volume integrals of $SA$ Eq. \eqref{e:sa:sa-ana} will be investigated.
\begin{eqnarray}
SA_v & = &\int_\Omega\int_\Omega\frac{1}{\gamma_a}\left[\frac{\rho_a^{2k}f_b+\rho_b^{2k}f_a}{\rho_a^k\rho_b^k}\uvec{B}_a + f_a\frac{\rho_a^k\rho_b^k}{\rho_a^{2k}}(\uvec{B}_b-\uvec{B}_a)\right] \cdot \unabla_a w_{ab} \uvec{\td r}_b \uvec{\td r}_a\\
& = & \int_\Omega\int_\Omega\frac{1}{\gamma_a}\left[\frac{\rho_a^k}{\rho_b^k}f_b\uvec{B}_a + \frac{\rho_b^k}{\rho_a^k}f_a\uvec{B}_b\right] \cdot \unabla_a w_{ab} \uvec{\td r}_b \uvec{\td r}_a\nonumber
\end{eqnarray}
Furthermore, due to the additivity of the integral we can split it up
\begin{eqnarray}
SA_v & = & \int_\Omega\int_\Omega\frac{1}{\gamma_a}\frac{\rho_a^k}{\rho_b^k}f_b\uvec{B}_a\cdot\unabla_a w_{ab} \uvec{\td r}_b \uvec{\td r}_a + \int_\Omega \int_\Omega \frac{1}{\gamma_a}\frac{\rho_b^k}{\rho_a^k}f_a\uvec{B}_b \cdot \unabla_a w_{ab} \uvec{\td r}_b \uvec{\td r}_a\\
& = & -\int_\Omega\int_\Omega\frac{1}{\gamma_a}\frac{\rho_a^k}{\rho_b^k}f_b\uvec{B}_a\cdot\unabla_b w_{ab} \uvec{\td r}_b \uvec{\td r}_a - \int_\Omega \int_\Omega \frac{1}{\gamma_a}\frac{\rho_b^k}{\rho_a^k}f_a\uvec{B}_b \cdot \unabla_b w_{ab} \uvec{\td r}_b \uvec{\td r}_a\nonumber.
\end{eqnarray}
In the last line the kernel gradient asymmetry $\unabla_a w_{ab} = - \unabla_b w_{ab}$ was used in both terms. The boundary part of Eq. \eqref{e:sa:sa-ana} can be reformulated to
\begin{eqnarray}
SA_b & = &\int_\Omega\int_{\partial\Omega}\frac{1}{\gamma_a}\left[-\frac{\rho_a^{2k}f_b+\rho_b^{2k}f_a}{\rho_a^k\rho_b^k}\uvec{B}_a + \frac{\rho_b^k}{\rho_a^k}f_a(\uvec{B}_a-\uvec{B}_b)\right]\cdot\uvec{n}_b w_{ab} \uvec{\td r}_b \uvec{\td r}_a\\
& = & \int_\Omega\int_{\partial \Omega}\frac{1}{\gamma_a}\left[-\frac{\rho_a^k}{\rho_b^k} f_b\uvec{B}_a - \frac{\rho_b^k}{\rho_a^k} f_a\uvec{B}_b\right] \cdot \uvec{n}_b  w_{ab} \uvec{\td r}_b \uvec{\td r}_a\nonumber\\
& = & -\int_\Omega\int_{\partial \Omega}\frac{1}{\gamma_a}\frac{\rho_a^k}{\rho_b^k} f_b\uvec{B}_a\cdot \uvec{n}_b w_{ab} \uvec{\td r}_b \uvec{\td r}_a - \int_\Omega\int_{\partial \Omega}\frac{1}{\gamma_a}\frac{\rho_b^k}{\rho_a^k} f_a\uvec{B}_b \cdot \uvec{n}_b  w_{ab} \uvec{\td r}_b \uvec{\td r}_a.\nonumber
\end{eqnarray}
Combining the volumic $SA_v$ and the boundary term $SA_b$ yields
\begin{eqnarray}
SA & = & SA_v + SA_b \\
&= &-\int_\Omega \int_\Omega\frac{1}{\gamma_a}\frac{\rho_a^k}{\rho_b^k}f_b\uvec{B}_a\cdot\unabla_b w_{ab}\uvec{\td r}_b \uvec{\td r}_a -\int_\Omega \int_{\partial \Omega}\frac{1}{\gamma_a}\frac{\rho_a^k}{\rho_b^k}f_b\uvec{B}_a\cdot\uvec{n}_b w_{ab}\uvec{\td r}_b \uvec{\td r}_a \nonumber\\
&& -\int_\Omega \int_\Omega\frac{1}{\gamma_a}\frac{\rho_b^k}{\rho_a^k}f_a\uvec{B}_b\cdot\unabla_b w_{ab}\uvec{\td r}_b \uvec{\td r}_a -\int_\Omega \int_{\partial \Omega}\frac{1}{\gamma_a}\frac{\rho_b^k}{\rho_a^k}f_a\uvec{B}_b\cdot\uvec{n}_b w_{ab}\uvec{\td r}_b \uvec{\td r}_a \nonumber\\
&= &-\int_\Omega \frac{1}{\gamma_a} \rho^k_a\uvec{B}_a\cdot\left[\int_\Omega\frac{f_b}{\rho_b^k}\unabla_b w_{ab}\uvec{\td r}_b - \int_{\partial \Omega}\frac{f_b}{\rho_b^k}\uvec{n}_b w_{ab}\uvec{\td r}_b\right] \uvec{\td r}_a \nonumber\\
&& -\int_\Omega\frac{1}{\gamma_a}\frac{f_a}{\rho_a^k} \left[\int_\Omega \rho_b^k\uvec{B}_b\cdot\unabla_b w_{ab}\uvec{\td r}_b -\int_{\partial \Omega}\rho_b^k\uvec{B}_b\cdot\uvec{n}_b w_{ab}\uvec{\td r}_b\right] \uvec{\td r}_a \nonumber\\
&= &-\int_\Omega \rho^k_a\uvec{B}_a\cdot\left[ \frac{1}{\gamma_a}\int_\Omega\unabla_b\left(\frac{f_b}{\rho_b^k}\right) w_{ab}\uvec{\td r}_b\right]\uvec{\td r}_a -\int_\Omega\frac{f_a}{\rho_a^k} \left[\frac{1}{\gamma_a}\int_\Omega \unabla_b\cdot\left(\rho_b^k\uvec{B}_b\right) w_{ab}\uvec{\td r}_b\right]\uvec{\td r}_a,\nonumber
\end{eqnarray}
where in the last step a reverse integration by parts was used. The terms in square brackets represent SPH approximations which in the limit of $h\rightarrow0$ converge to
\begin{eqnarray}
SA & \rightarrow & - \int_\Omega \rho_a^k\uvec{B}_a\cdot\unabla_a\left(\frac{f_a}{\rho_a^k}\right)\uvec{\td r}_a - \int_\Omega\frac{f_a}{\rho_a^k}\unabla_a\cdot\left(\rho_a^k\uvec{B}_a\right)\uvec{\td r}_a\\
& = & - \int_\Omega \left[\rho_a^k\uvec{B}_a\cdot\unabla_a\left(\frac{f_a}{\rho_a^k}\right)-\frac{f_a}{\rho_a^k}\unabla_a\cdot\left(\rho_a^k\uvec{B}_a\right)\right]\uvec{\td r}_a\nonumber\\
& = & - \int_\Omega \unabla_a\cdot\left(f_a\uvec{B}_a\right)\uvec{\td r}_a\nonumber\\
& = & - \int_{\partial\Omega}f_a\uvec{B}_a\cdot\uvec{n}_a\uvec{\td r}_a,\nonumber
\end{eqnarray}
where in the second last line again a reverse integration by parts was used and the final line follows from Stokes' theorem.

\end{document}